\documentclass[lettersize,journal]{IEEEtran}

\usepackage{amsmath,amsfonts,amssymb}
\usepackage{algorithmic}
\usepackage{algorithm}
\usepackage{array}
\usepackage[caption=false,font=normalsize,labelfont=sf,textfont=sf]{subfig}
\usepackage{textcomp}
\usepackage{stfloats}
\usepackage{url}
\usepackage{verbatim}
\usepackage{graphicx}
\usepackage{cite}
\usepackage{epstopdf}
\usepackage{xcolor}
\usepackage{multirow}

\newtheorem{theorem}{Theorem}

\begin{document}

\title{Joint Mobility Control and MEC Offloading for Hybrid Satellite-Terrestrial-Network-Enabled Robots}

\author{Peng~Wei, Wei~Feng,~\IEEEmembership{Senior Member, IEEE}, Yanmin~Wang, Yunfei~Chen,~\IEEEmembership{Senior Member, IEEE}, Ning~Ge,~\IEEEmembership{Member, IEEE}, and Cheng-Xiang~Wang,~\IEEEmembership{Fellow, IEEE}

\thanks{
	The work of Peng Wei, Wei Feng, and Ning Ge was supported by the National Key Research and Development Program of China under Grant 2020YFA0711301, the National Natural Science Foundation of China under Grants U22A2002 and 61901298, and the Suzhou Science and Technology Project.
	The work of Yunfei Chen was supported by the King Abdullah University of Science and Technology Research Funding (KRF) under Award ORA-2021-CRG10-4696.
	The work of Cheng-Xiang Wang was supported by the National Natural Science Foundation of China under Grant 61960206006, the Key Technologies R\&D Program of Jiangsu (Prospective and Key Technologies for Industry) under Grants BE2022067 and BE2022067-1, and the EU H2020 RISE TESTBED2 project under Grant 872172.
	Part of this paper has been accepted by the IEEE ICC 2023. 
	{\it(Corresponding author: Wei Feng.)}

Peng Wei, Wei Feng, and Ning Ge are with the Department of Electronic Engineering, Beijing National Research Center for Information Science and Technology, Tsinghua University, Beijing 100084, China (e-mail: wpwwwhttp@163.com; fengwei@tsinghua.edu.cn; gening@tsinghua.edu.cn).

Yanmin Wang is with the School of Information Engineering, Minzu University of China, Beijing 100041, China (e-mail: yanmin-226@163.com).

Yunfei Chen is with the Department of Engineering, University of Durham, Durham DH1 3LE, U.K. (e-mail: yunfei.chen@durham.ac.uk).

Cheng-Xiang Wang is with the National Mobile Communications Research Laboratory, School of Information Science and Engineering, Southeast University, Nanjing 210096, China, and also with the Purple Mountain Laboratories, Nanjing 211111, China (e-mail: chxwang@seu.edu.cn).
}
}

\maketitle

\begin{abstract}
	Benefiting from the fusion of communication and intelligent technologies, network-enabled robots have become important to support future machine-assisted and unmanned applications.
	To provide high-quality services for robots in wide areas, hybrid satellite-terrestrial networks are a key technology.
	Through hybrid networks, computation-intensive and latency-sensitive tasks can be offloaded to mobile edge computing (MEC) servers.
	However, due to the mobility of mobile robots and unreliable wireless network environments, excessive local computations and frequent service migrations may significantly increase the service delay.
	To address this issue, this paper aims to minimize the average task completion time for MEC-based offloading initiated by satellite-terrestrial-network-enabled robots.
	Different from conventional mobility-aware schemes, the proposed scheme makes the offloading decision by jointly considering the mobility control of robots.
	A joint optimization problem of task offloading and velocity control is formulated.
	Using Lyapunov optimization, the original optimization is decomposed into a velocity control subproblem and a task offloading subproblem.
	Then, based on the Markov decision process (MDP), a dual-agent reinforcement learning (RL) algorithm is proposed.
	The convergence and complexity of the improved RL algorithm are theoretically analyzed, and the simulation results show that the proposed scheme can effectively reduce the offloading delay.
\end{abstract}


\section{Introduction}

With the rapid development of communication and intelligent technologies, network-enabled robots have become an important application for the advancement of the future society, such as in assisted-living, industry, and transport environments \cite{8166737, AllianceReport, You2020,Buchong1,Buchong2}.
When robots operate in wide areas, the hybrid satellite-terrestrial network is key to provide ubiquitous coverage and information perception \cite{9453860,9183767}.
Since robots, especially mobile robots, always have limited computing capabilities and storage capacities, their computation-intensive and delay-sensitive tasks can be uploaded to powerful edge servers with the aid of hybrid networks.
This is known as task offloading in mobile edge computing (MEC) \cite{7879258}.
Meanwhile, to efficiently complete specific missions within a given time, such as in cooperation among multiple robots, these robots also need to perform some constrained movements.
However, due to the mobility of robots and time-varying requirements of task offloading, hybrid satellite-terrestrial networks are highly dynamic. 
A service migration occurs when a robot moves away from its original location, and thus, its current MEC server that provides mobile service is different from the previous MEC server \cite{8000803}.
Furthermore, compared with conventional clouds, MEC systems have limited computation and storage resources \cite{7879258}.
On the other hand, wireless environments are unreliable \cite{LIU202296,9786750,Buchong5,Buchong6,Buchong7,6362134}. 
In this regard, when a large number of mobile robots access the hybrid network, frequent service migrations may deteriorate the hybrid network environment, such as network overload and packet loss.
As a result, the service delay of offloaded tasks for hybrid satellite-terrestrial-network-enabled robots will be significantly increased.

Many methods have been proposed to address the migration problem in MEC-based terrestrial networks \cite{9305956,9174795,9014294,8959146,9416470,8924682}.
In \cite{9305956}, considering distributed user mobility, a multi-agent reinforcement learning (RL) algorithm was presented. 
To minimize the task offloading delay with the accumulated service migration cost, the MEC-based digital twin network was optimized by RL in \cite{9174795}.
To balance the high quality of services (QoS) and migration cost, \cite{9014294} proposed a deep RL enabled optimization scheme in a vehicular network.
Furthermore, according to predictable trajectories and mobility-induced communication rates, a mobility-aware task offloading policy was designed in \cite{8959146}.
By assigning velocity-based access priorities to mobile devices \cite{9416470}, speed-aware task offloading was optimized by RL.
Leveraging mobility, \cite{8924682} devised a microservice coordination scheme to minimize the overall service delay. 
However, when the satellite communication is incorporated, the heterogeneity between satellite and cellular communication systems, such as different propagation delays and different communication rates, may result in higher service latencies.

Thus, an increasing number of studies on MEC based on satellite-terrestrial networks have been conducted for cooperative offloading \cite{8610431,9685309,8689236,8689224,9043505} and service migrations \cite{9139102,9351537,9758064,9685350}.
In \cite{8610431}, a cooperative computation offloading model was designed to provide high-speed services.
In \cite{9685309}, to minimize energy consumption in computation offloading, a cloud-edge collaboration problem was optimized by RL and successive convex approximation algorithms. 
By considering user preference and evolved satellite processing capabilities, \cite{8689236} proposed satellite-terrestrial cooperation-based double-edge networks to relieve terrestrial backhaul burdens.
To efficiently allocate the distributed MEC servers, the joint optimization of energy consumption and delay was considered in double-edge networks \cite{8689224}.
To jointly optimize the user association, resource allocation, and offloading policy in Internet of Things (IoT) networks using multiple satellites and their gateways, the cost of delay and energy consumption was minimized by the Lagrange multiplier and RL algorithms \cite{9043505}.
In addition, to reduce the migration cost, a service migration model was devised in \cite{9139102} based on task characteristics to make a tradeoff between task completion time and energy consumption.
In \cite{9351537}, the live migration of a virtual network function (VNF) with its reinstantiation and scheduling was studied.
Then, two Tabu search-based algorithms were employed for dynamic VNF mapping and scheduling.
For low-delay airplane applications in \cite{9758064}, the in-flight service provisioning problem was formulated by routing and reconfiguration and solved by the online heuristic algorithm.
Furthermore, a distributed two-layer decomposition model was proposed to minimize the migration cost in \cite{9685350}.
These existing works consider how to optimize task offloading based on the mobility of devices, but they do not consider how to optimize task offloading based on mobility control.

Even in MEC-based terrestrial networks, most mobility-related network optimization works for task processing and resource allocation are from the supply side that involve resource scheduling \cite{9366426}; wireless control \cite{7749210,9322156}; and task offloading \cite{9552188,9284254}.
In these methods, the use of mobility includes mobility prediction \cite{9366426}; mobility state sharing \cite{7749210}; mobility control and its stability \cite{9322156}; and velocity-based task classification \cite{9552188,9284254}.
Additionally, in \cite{Buchong9}, Wu {\it et al.} studied computation offloading in multi-access MEC to minimize the overall offloading delay of mobile users.
In \cite{Buchong8}, using wireless power transfer for mobile devices with limited energy capacities, a joint optimization of total energy consumption and the learning convergence latency was proposed. 
Although the effect of mobility control has been analyzed in \cite{9322156}, task offloading and service migration in hybrid satellite-terrestrial networks have not been considered, and the case of no wireless network coverage caused by damage to network infrastructures or heavy network loads is not included either.

Different from conventional network optimization approaches \cite{9366426,7749210,9322156,9552188,9284254,Buchong9,Buchong8}, a demand shaping-based approach was designed from the user side in \cite{8506622,6736762,6082509, 5679756}.
Based on the willingness of users to move, a closed-loop system model for spatial control and temporal control was developed. 
In spatial control, users are encouraged to move from a severely congested location to a less congested location.
In temporal control, an incentive design for reducing the data demand of users in a severely congested location is proposed.
However, these approaches are intended for humans, not robots.
Furthermore, when all wireless channels are unavailable, mobility control for mobile robots with offloading requirements is not considered.
The channel unavailability might be due to a large number of mobile robots accessing limited communication resources, severe channel fading, damage to the access point (AP), and so on.
In this case, in the AP coverage where all channels are unavailable, the low-speed movements of mobile robots increase local computations.
When wireless channels are available in the AP coverage, high-speed movements of mobile robots lead to more service migrations.
As a result, the total service delay is significantly increased.

Motivated by the above observations, in this work, we develop a joint velocity control and MEC-based offloading strategy to improve the QoS in hybrid satellite-terrestrial networks.
We consider a scenario where multiple mobile robots cooperate with each other to complete the assigned mission.
At the same time, they periodically sense their surroundings, offload data to cellular/satellite MEC servers for processing, and receive computational results from selected MEC servers. 
Among the surrounding information, the availability of wireless communication is the crucial state information for velocity control, where wireless communication refers to radio communication in this paper.
Our objective is to minimize the average task completion time for MEC-based offloading when a mobile robot travels an entire trip.
We formulate the long-term offloading problem as a Markov decision process (MDP) problem and propose a joint optimization for velocity control and task offloading.
Our main contributions are summarized as follows.
\begin{itemize}
	\item A joint optimization problem of task offloading, velocity control, and service migration is proposed in this paper, which is formulated to minimize the average task completion time.
	The non-deterministic polynomial-time hardness (NP-hardness) of the optimization problem is proven.
	Then, the relationship between the velocity control and task offloading is explored.
	\item According to the coverage regions of APs, a Lyapunov optimization-based decomposition is employed, which can achieve task offloading optimization and velocity control optimization.
	Then, based on the MDP, a dual-agent RL algorithm is employed to obtain the effective decision-making of task offloading and velocity control.
	Furthermore, the convergence and complexity of the improved RL algorithm are analyzed in terms of $Q$ functions.
	Finally, simulation results show that compared with conventional offloading schemes, the proposed scheme can save up to 40\% of the average task completion time.
\end{itemize}

The rest of this paper is organized as follows. 
In Section II, the system model of task offloading and velocity control for satellite-terrestrial-network-enabled robots is introduced. 
In Section III, the optimization model and its NP-hardness are given, and the dual-agent RL algorithm and its convergence are proposed.
In Section IV, simulation results are provided.
Finally, in Section V, conclusions are drawn.

\section{System Model}

\subsection{System Description}

\begin{figure*}[t]
	\setlength{\abovecaptionskip}{0.cm}
	\setlength{\belowcaptionskip}{-0.cm}
	\centerline{\includegraphics[width=4.85in]{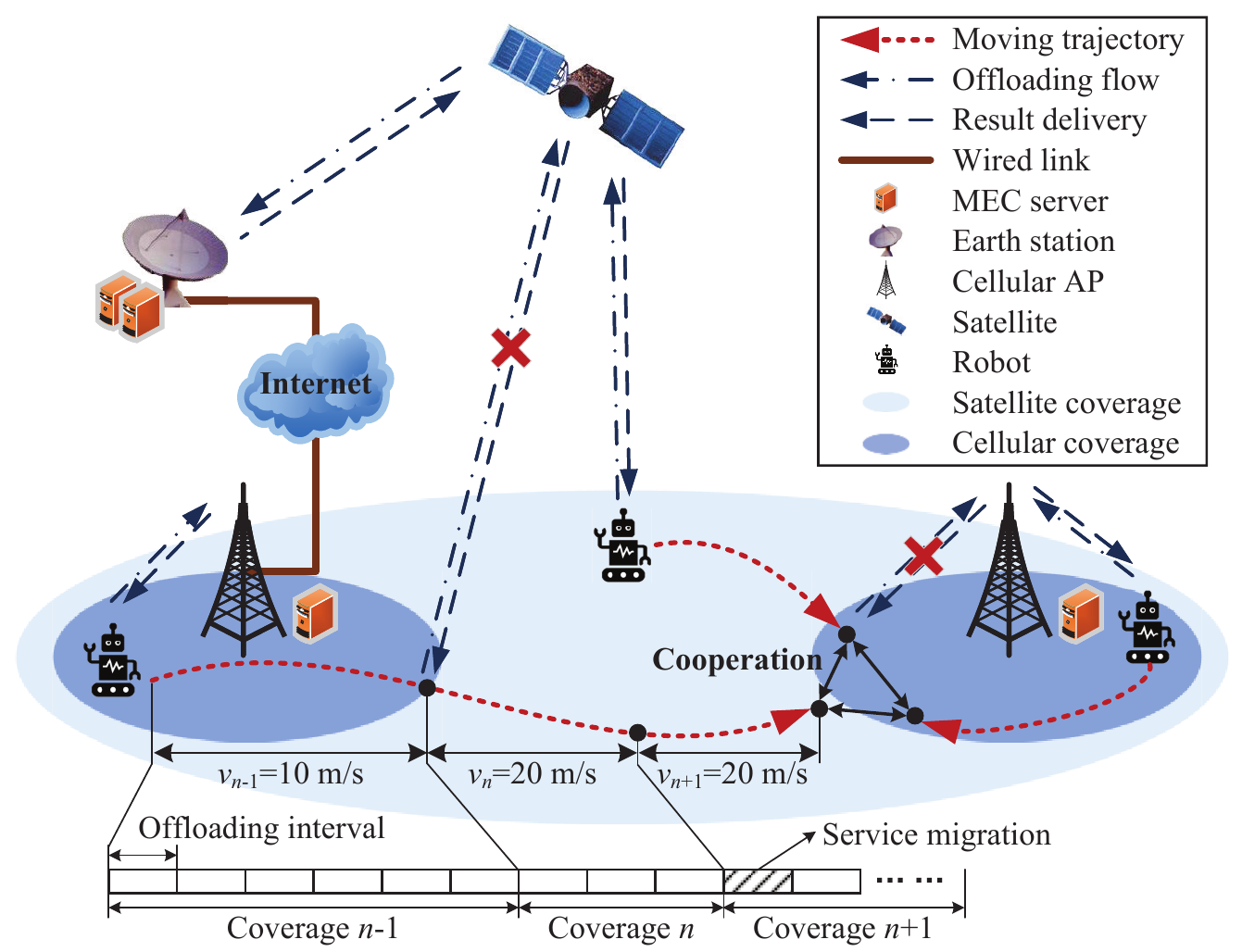}}
	\caption{System model for multiple robots in the hybrid satellite-terrestrial-network with MEC, where $v_n$ denotes the average moving velocity in the $n$th AP coverage region.}
	\label{Fig:Fig1}
\end{figure*}

As shown in Fig. \ref{Fig:Fig1}, a hybrid satellite-terrestrial network with MEC is considered, which can enable multiple robots.
Through the hybrid network, these robots intelligently cooperate to complete a specific mission within a given time.
To avoid collisions and to complete the collaborative mission on time, efficient control of robot mobility is needed, such as velocity control.
Using appropriate velocity control, these robots can move to their designated positions to complete the collaborative mission in time. 
During the movement, mobile robots also need to complete other tasks to make autonomous decisions.
For example, to handle computation-intensive or latency-sensitive tasks, they periodically offload these tasks (such as sensor data) to and receive the computational results from MEC servers through APs.
Thus, these robots are multitasking robots that not only complete mobility-controlled cooperative missions but also execute MEC-empowered offloading.
For the multi-mobile robot scenario, we focus on a mobile robot moving from one place to another place within a given time in a complex network environment. 
The reasons include the following.
1) The joint optimization of velocity control and MEC-powered offloading for a satellite-terrestrial-network-enabled robot has not been studied in existing works.
Since joint optimization involves two different systems, a hybrid network system and a motion control system, it is also a complex optimization problem.
2) The autonomous decision-making behaviors and mobility behaviors of other robots may complicate the network environment, such as generating the dynamic network load and intermittent communication environment.
Hence, other robot-induced network dynamics are regarded as an important part of the complex network environment.
A severely degraded network environment will significantly affect data uploading and result downloading.
As a result, the velocity control and task offloading decisions of the selected mobile robot cannot be executed in time.

The availability of wireless communication is also included in the periodically perceived surrounding information of the mobile robot. 
When wireless communication is available, periodic offloading is executed by the mobile robot.
Each period is referred to as an offloading interval $\Delta T$.
The wireless communication system has $N$ APs, including $N_1$ cellular APs and $N_2$ satellites, i.e., $N=N_1+N_2$.
When wireless communication is unavailable, the mobile robot can process tasks using its limited computing capability.
When the velocity of the robot is too low, it cannot reach the destination on time and complete the cooperative mission with other robots.
Thus, an average velocity should be maintained by the appropriate velocity control to guarantee that the moving time of the mobile robot is smaller than or equal to the due time $T_{\rm move}$ in the whole trip.
To coordinate offloading and mobility, we focus on the joint optimization of task offloading and velocity control as a complement to conventional offloading approaches.

In the sequel, we focus on cellular AP-based MEC servers and satellite-based MEC servers, indexed by $\mathcal{N}_1$ and $\mathcal{N}_2$ $(\mathcal{N}_1 \cup \mathcal{N}_2=\{1,2,\ldots,N\})$, respectively.
For each cellular AP/Earth station, an independent MEC server is equipped, where the MEC server equipped with the Earth station has more computing capabilities than the MEC server equipped with the cellular AP.
The MEC server receives and computes the offloaded tasks, and sends the computational results back to the mobile robot.
Optical connections are assumed between MEC servers, between APs, and between APs and MEC servers.
It is noted that the satellite MEC servers and cellular MEC servers are connected through the Internet.

In each AP coverage region, when the mobile robot is assumed to move a fixed distance, the number of offloading intervals depends on its moving velocity. 
A low velocity increases the offloading intervals as well as the moving time.
As a result, the mobile robot may not be able to reach the destination on time.
More importantly, when wireless communication is unavailable in the current AP coverage region, more computational tasks are undertaken by the selected mobile robot. 
Since the mobile robot has much smaller computational resources than the MEC server, the task completion time will be significantly increased.
On the other hand, a high velocity increases the handover of the mobile robot between multiple APs.
In this case, we assume that the information of all robots (in terms of locations and offloading requests) served by different MEC servers is available \cite{9509399}.
Based on such global information, the incomplete task in the current MEC server may be migrated to another MEC server closest to the mobile robot.
As a result, the migration time is increased \cite{9509399, 6638730, 9174795}.
Therefore, due to the uncertainty of wireless link availability, inappropriate velocities could increase the task completion time in MEC-based offloading.

An example of task offloading and velocity control is shown in Fig. \ref{Fig:Fig1}. 
A mobile robot periodically generates tasks (e.g., in each offloading interval).
In the $(n-1)$th coverage region, the mobile robot with $v_{n-1}=10$ m/s has twice as many offloading intervals as the mobile robot with $v_{n}=20$ m/s.
As a result, when wireless communication is unavailable in the $(n-1)$th AP, a slower movement will further increase the offloading intervals and the number of local computations.
In contrast, in the $n$th AP with available wireless communication, the rapid movement reduces the dwell time of the mobile robot.
In this case, some tasks offloaded to the $n$th MEC server may not be fully processed, and their computational results cannot be returned by the $n$th AP.
Based on the global information of the robot-located coverage region and offloading request, the uncompleted tasks can be migrated from the $n$th MEC server to the $(n+1)$th MEC server, and their computational results will be sent back by the $(n+1)$th AP.
As a result, the rapid movement increases the number of service migrations.
According to \cite{9552188}, the low velocity has a high delay tolerance, and the high velocity has a low delay tolerance.
Thus, considering the velocity control of the mobile robot and wireless communication availability, we focus on minimizing the process-oriented average task completion time in MEC-based offloading using the hybrid satellite-terrestrial network.

\subsection{Offloading Model}

For the task offloading of the mobile robot, the local computation, wireless communication, MEC computation, and service migration are described below. 

\subsubsection{Local computation model}
When wireless communication is unavailable in the current AP, the task is computed by the mobile robot.
The local computation delay $T_{\rm local}(t)$ in the $t$th interval is expressed as
\begin{equation}\label{eq:eq1}
T_{\rm local}(t)=\left(1-\textbf{1}_{\alpha}(m,t)\right) \frac{D(t)\Phi}{f_{\rm local}(t)},
\end{equation}  
where the indicator function $\textbf{1}_{\alpha}(m,t)$ stands for the offloading decision in the $t$th interval, $\textbf{1}_{\alpha}(m,t)=1$ for $m \in \alpha=\{1,2,\ldots,N\}$ denotes that all data are offloaded to the $m$th MEC server in the $t$th interval, $\textbf{1}_{\alpha}(m,t)=0$ for $m=0$ indicates the local computation, $D(t)$ is the data size generated by the mobile robot in the $t$th interval, $\Phi$ is the required CPU cycles per bit, and $f_{\rm local}(t)$ denotes the CPU frequency at the mobile robot. 

\subsubsection{Communication model}
We assume that the same communication rate exists in the uplink and downlink.
When wireless communication is available in the current cellular AP, the task generated from the mobile robot can be offloaded to an MEC server over wireless channels.
In the $t$th interval, the communication delay $T_{\rm com}(t)$ is expressed as
\begin{equation}\label{eq:eq2}
T_{\rm com}(t)=\frac{\textbf{1}_{\alpha}(m,t) \left(D(t) + \bar{D}(t) \right)}{W \log_2\left(1+\frac{p h^2}{\sigma^2}\right)},
\end{equation}
where $\bar{D}(t)$ is the data size of the computational results, $p$ is the transmit power, and $h$, $W$, and $\sigma^2$ denote the channel gain, bandwidth, and noise power, respectively. 

When the offloaded task is transferred from the satellite to the Earth station, the communication delay $T_{\rm com}(t)$ is 
\begin{align}\label{eq:eq3}
T_{\rm com}(t) \!= \! \textbf{1}_{\alpha}(m,t) \! \bigg(\!\!  2 \frac{d_{\rm GS}\!+\!d_{\rm SE}}{c} 
 \!+\! \left( D(t) \!+\! \bar{D}(t) \right) \!\! \left(\! \frac{1}{r_{\rm GS}} \!+\! \frac{1}{r_{\rm SE}}\! \right)\!\!\! \bigg)\!,
\end{align}
where $d_{\rm GS}$ and $d_{\rm SE}$ denote the distances from the mobile robot to the satellite and from the satellite to the Earth station, respectively, $c$ is the speed of light, and $r_{\rm GS} = W_{\rm GS} \log_2\left(1+\frac{p_{\rm GS} h^2_{\rm GS}}{\sigma^2_{\rm GS}}\right)$ and $r_{\rm SE} = W_{\rm SE} \log_2\left(1+\frac{p_{\rm SE} h^2_{\rm SE}}{\sigma^2_{\rm SE}}\right)$ denote the communication rates in the links between the mobile robot and the satellite and between the satellite and the Earth station, respectively.
$W_{\rm GS}$ (or $W_{\rm SE}$), $p_{\rm GS}$ (or $p_{\rm SE}$), $h_{\rm GS}$ (or $h_{\rm SE}$), and $\sigma^2_{\rm GS}$ (or $\sigma^2_{\rm SE}$) stand for bandwidth, transmit power, channel gain, and noise power in robot-satellite (or satellite-Earth station) transmission. 

\subsubsection{MEC model}
With the aid of wireless transmission in the current AP, when the task is offloaded to the $m$th MEC server in the $t$th interval, the computation delay $T_{\rm MEC}(t)$ is given by
\begin{equation}\label{eq:eq4}
T_{\rm MEC}(t)=\frac{\textbf{1}_{\alpha}(m,t) D(t) \Phi}{f_{{\rm MEC},m}(t)},
\end{equation}
where $f_{{\rm MEC},m}(t)$ is the CPU frequency of the $m$th MEC server.

\subsubsection{Migration model}
Due to the mobility of the robot, the corresponding service provider (e.g., virtual machine (VM)) is migrated from the initial MEC server to the current counterpart through one- or multi-hop optical communications.
In this case, the service downtime may cause a delay that cannot be ignored. 
In our model, when the MEC server in the $(t-1)$th interval is different from the MEC server in the $t$th interval, a migration delay will be incurred, which is expressed as
\begin{align}\label{eq:eq5}
T_{\rm mig}(t)
=& I \big\{ {\rm M}(t \!-\! 1) \neq {\rm M}(t) \cap \left( {\rm M}(t \!-\! 1) \neq 0 \cup {\rm M}(t) \neq 0 \right)  \nonumber \\
& \cap \left( {\rm M}(t \!-\! 1) \in \mathcal{N}_1 \cup {\rm M}(t) \in \mathcal{N}_1 \right) \big\} C  \nonumber \\
&+ I \big\{ \left( {\rm M}(t \!-\! 1) \in \mathcal{N}_1 \cap {\rm M}(t) \in \mathcal{N}_2 \right)  \nonumber \\
 & \cup \left( {\rm M}(t \!-\! 1) \in \mathcal{N}_2 \cap {\rm M}(t) \in \mathcal{N}_1 \right) \big\} \Delta C,
\end{align}
where ${\rm M}(t)\in \mathcal{N}_1 \cup \mathcal{N}_2 \cup \{0\}$, ${\rm M}(t)=0$ denotes the local computing, ${\rm M}(t)=1,2,\ldots,N$ denotes the MEC server, and $I\{\cdot\}=1$ if the condition in $\{\cdot\}$ is satisfied and otherwise, $I\{\cdot\}=0$.
$C$ denotes the migration time cost, which is replaced by $\rho T_{\rm MEC}(t)$ with the scaling factor $\rho$ ($\rho \in [0,1]$) in this paper. 
$\Delta C$ is the extra migration cost between the cellular MEC server and the satellite-based MEC server.

\subsection{Velocity Control Model}
We first assume that the robot has a preplanned timestamped reference trajectory to reduce its mobility model to one-dimensional motion, that is, velocity control without considering direction.
Then, we assume that $v_n(l_0)$ and $v_n(l_{\rm end})$ are the initial velocity and final velocity in the $n$th AP coverage region $(n=1,2,\ldots,N)$, respectively. 
We also assume $v_n(l_{\rm end})=v_{{\rm goal},n}$ with the target velocity $v_{{\rm goal},n} \in [v_{\rm min}, v_{\rm max}]$ and a given acceleration $a>0$.
Note that the target velocity $v_{{\rm goal},n}$ is not known in advance and needs to be solved in the joint optimization problem in the next section.
When the mobile robot enters the $n$th AP coverage region, its velocity control policy should be first determined by the relationship between $v_n(l_0)$ and $v_{{\rm goal},n}$. 
Thus, the instantaneous velocity $v_n(l)$ in the $l$th interval of the $n$th AP coverage region can be expressed as
\begin{equation}\label{eq:eq7}
	v_n(l) \!=\!
	\begin{cases}
		\min\left\{v_n(l_0) + a l,  v_{{\rm goal},n} \right\}\!,\!\!   & v_n(l_0)<v_{{\rm goal},n}, \\
		v_n(l_0),   &  v_n(l_0)=v_{{\rm goal},n}, \\
		\max\left\{v_n(l_0) - a l, v_{{\rm goal},n} \right\}\!,\!\!  &  v_n(l_0)>v_{{\rm goal},n},
	\end{cases}
\end{equation}
where $v_n(l_0)<v_{{\rm goal},n}$, $v_n(l_0)=v_{{\rm goal},n}$, and $v_n(l_0)>v_{{\rm goal},n}$ correspond to the velocity controls of acceleration, constant movement, and deceleration, $l \in \mathcal{L}_n = \{1,2,\ldots,L_n\}$, and $L_n$ is the number of offloading intervals in the $n$th AP coverage region, which is calculated by
\begin{equation}\label{eq:eq10}
	L_n= \left\lfloor \frac{T_{{\rm goal},n}}{\Delta T}  \right\rfloor,
\end{equation}
where $T_{{\rm goal},n}$ is the travel time across the $n$th AP coverage region, given as 
\begin{equation}\label{eq:eq11}
	T_{{\rm goal},n} \!=\! 
	\begin{cases}
	\frac{1}{v_{{\rm goal},n}}  \left( c_n \!+\! \frac{\left( v_{{\rm goal},n}-v_n(l_0) \right)^2}{2a} \right)\!,    & v_n(l_0) \leqslant  v_{{\rm goal},n}, \\
		\frac{1}{v_{{\rm goal},n}}  \left( c_n \!-\! \frac{\left( v_n(l_0) - v_{{\rm goal},n} \right)^2}{2a} \right)\!,    & v_n(l_0)  >  v_{{\rm goal},n},
	\end{cases}
\end{equation}
where $c_n$ is the moving distance in the $n$th AP coverage region.

Therefore, when the mobile robot passes through the $n$th AP, for $t=1,2,\ldots,\sum^{N}_{n=1}L_n$, the task completion time $T_n(t)$ in the $t$th interval can be formulated as
\begin{equation}\label{eq:eq8}
T_n(t) \!=\! T_{\rm local}(t) \!+\! \mu_n \left( T_{\rm com}(t) \!+\! T_{\rm MEC}(t) \!+\! T_{\rm mig}(t)\right)\!,
\end{equation}
where $\mu_n=0$ and $1$ denote the states when all wireless channels in the $n$th AP are unavailable or available, respectively.
In the entire journey, the average task completion time is expressed as 
\begin{equation}\label{eq:eq9}
T_{\rm mean}= \frac{1}{\sum\limits^{N}_{n=1}L_n} \sum\limits^{N}_{n=1} \sum\limits^{L_n}_{l=1} T_n(l) .
\end{equation}

\section{Joint Mobility Control and MEC Offloading}

\subsection{Optimization Problem}
To enhance the QoS for delay-sensitive applications, the optimization problem is formulated as 
\begin{subequations} \label{eq:eq12}
\begin{align}
\min\limits_{v_{{\rm goal},n}, \textbf{1}_{\alpha}(m,t)} & T_{\rm mean} \label{eq:eq12a} \\
\text{s.t.}~~
& T_n(t) \leqslant T_{n,{\rm max}}(t) \label{eq:eq12b} \\
&\sum\limits^{N}_{n=1} T_{{\rm goal},n} \leqslant T_{\rm move}  \label{eq:eq12c} \\
& \sum\limits^{N}_{m=1} \textbf{1}_{\alpha}(m,t) \leqslant 1  \label{eq:eq12d} \\
&v_{{\rm goal},n} \in [v_{\rm min}, v_{\rm max}]  \label{eq:eq12e} \\
&\textbf{1}_{\alpha}(m,t) \in \{0,1\}  \label{eq:eq12f}
\end{align}
\end{subequations}
Constraint \eqref{eq:eq12b} indicates that the task completion time cannot be larger than the maximal completion time, where $T_{n,{\rm max}}(t)$ is the computation time for all data to be computed locally.  
Constraint \eqref{eq:eq12c} denotes the tolerable total moving time across the whole journey.
Constraint \eqref{eq:eq12d} guarantees that only one MEC server is selected or only one local computation is performed per offloading interval.
Constraints \eqref{eq:eq12e} and \eqref{eq:eq12f} involve the decisions of the velocity control and task offloading.

\begin{theorem}\label{Theo:Theo1}
The problem in \eqref{eq:eq12} is NP-hard.
\end{theorem}

\begin{IEEEproof}
	See Appendix \ref{App:App1}.	
\end{IEEEproof}

\begin{figure*}[t]
	\setlength{\abovecaptionskip}{0.cm}
	\setlength{\belowcaptionskip}{-0.cm}
	\centerline{\includegraphics[width=5.1in]{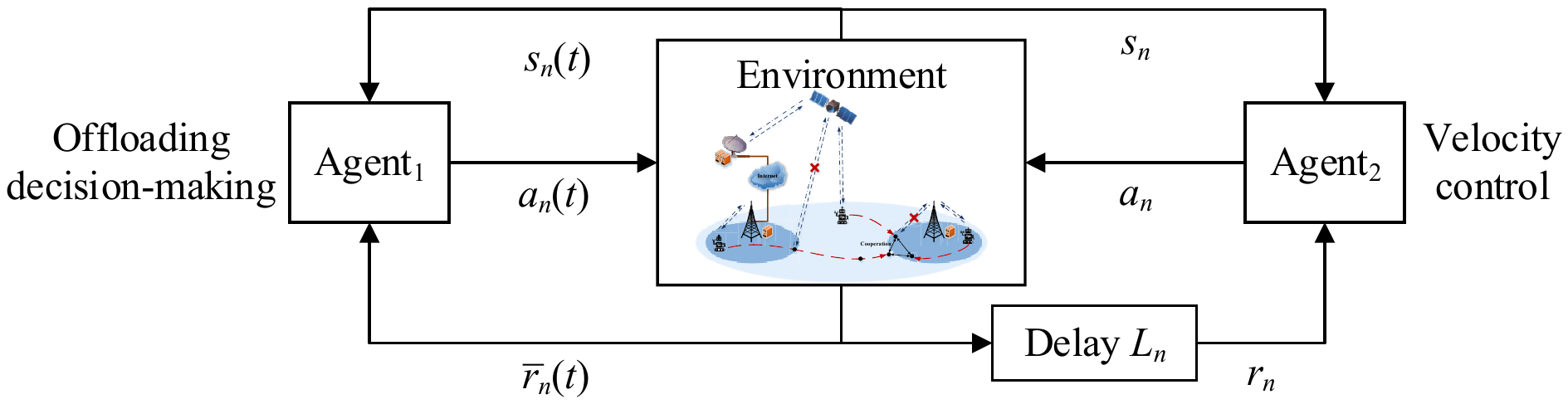}}
	\caption{The framework of dual-agent $Q$-learning.}
	\label{Fig:Fig2}
\end{figure*}

Furthermore, for the process-oriented optimization problem \eqref{eq:eq12}, due to the robot's mobility, time-varying wireless channel, and dynamic computing capability, conventional optimization methods are computationally inefficient.
Machine learning is a promising alternative \cite{Buchong3,Buchong4}, such as $Q$-learning. 
However, the conventional $Q$-learning method cannot be directly employed to solve Problem \eqref{eq:eq12}.
The first reason is the curse of dimensionality caused by the large-scale state space and action space.
The second reason involves asynchronous actions, where the offloading decision is updated per offloading interval, and the velocity control decision is made per AP coverage region.
Thus, based on Lyapunov optimization \cite{9174795}, decomposing the optimization problem \eqref{eq:eq12} over the whole journey into multiple subproblems over each AP coverage region yields 
\begin{subequations} \label{eq:eq13}
\begin{align}
\min\limits_{v_{{\rm goal},n}, \alpha_m(l)} & \frac{1}{L_n} \sum\limits^{L_n}_{l=1} T_n(l) \label{eq:eq13a} \\
\text{s.t.}~~
& T_{{\rm goal},n} \leqslant k_n T_{\rm move} \label{eq:eq13b} \\
& \eqref{eq:eq12b}, \eqref{eq:eq12d}-\eqref{eq:eq12f}  \label{eq:eq13c}
\end{align}
\end{subequations}
where $k_n={c_n}/{\sum^{N}_{n=1}c_n}$.

\begin{theorem} \label{Theo:Theo2}
Problem \eqref{eq:eq13} is NP-hard.
\end{theorem}

\begin{IEEEproof}
See Appendix \ref{App:App2}.	
\end{IEEEproof}

\subsection{Optimization Based on Dual-Agent $Q$-Learning}
In this subsection, as shown in Fig. \ref{Fig:Fig2}, a general framework of dual-agent $Q$-learning is provided to optimize the decision-making of offloading and velocity control in \eqref{eq:eq13}. 
In the $n$th AP coverage region, ${\rm Agent}_1$ makes the offloading decision.
Via accumulating the offloading reward and observing the channel state, the target velocity is determined by ${\rm Agent}_2$.
It is implied that ${\rm Agent}_1$ is the local agent, and ${\rm Agent}_2$ is the global agent. 

Based on the MDP, the states, actions, and rewards for offloading-related $Q$-learning and mobility-related $Q$-learning are modeled below.

In offloading-based $Q$-learning, the state includes: the current AP coverage region ${\rm AP}(t)$, current channel state $\mu_n$, data size $D(t)$, CPU frequency of the mobile robot $f_{\rm local}(t)$, current velocity $v_n(t)$, and previous MEC server ${\rm M}(t \!-\! 1)$, given as 
\begin{equation}\label{eq:eq14}
s_n(t)\!=\!\left\{{\rm AP}(t),  \mu_n, D(t), f_{\rm local}(t), v_n(t), {\rm M}(t \!-\! 1)\right\}\!.
\end{equation}
 The action includes the offloaded data ratio and current MEC server, which is expressed as
\begin{equation}\label{eq:eq15}
a_n(t)=\left\{\textbf{1}_{\alpha}(m,t) \right\}.
\end{equation}
In the reward design, the same reward is obtained in the velocity control when all possible target velocities $v_{{\rm goal},n}$ satisfy the moving constraint \eqref{eq:eq13b}.
Thus, the instantaneous reward for the velocity control is designed as $\max\left\{ \frac{1}{L_n}\left( T_{{\rm goal},n} \!-\! k_n T_{\rm move}\right), 0\right\}$. 
According to \eqref{eq:eq10}, this reward can be simplified as $\max\left\{ \Delta T \!-\! \frac{k_n}{L_n} T_{\rm move}, 0\right\}$.
By combining the offloading reward and velocity control reward, the instantaneous reward $\bar{r}_n(t)$ is given as 
\begin{align}\label{eq:eq16}
\bar{r}_n(t)=
 & (1-\theta) \exp\left( 1- \frac{T_n(t)}{T_{n,{\rm max}}(t)}\right) \nonumber \\
 & + \theta \exp \left( 1- \frac{\max\left\{ \Delta T - \frac{k_n}{L_n} T_{\rm move}, 0\right\}}{\frac{k_n}{L_n} \left(T_{\rm low} -T_{\rm move} \right)}\right),
\end{align}
where $\theta$ is the preference factor between offloading and velocity control, $T_{\rm low} =\sum^{N}_{n=1}c_n/v_{\rm min}$.
Finally, when legal action is obtained in the training, the reward in \eqref{eq:eq16} will be employed. 
On the contrary, when an illegal action occurs, the reward value is set to -1. 
Thus, in the training, the instantaneous reward is expressed as 
\begin{equation}\label{eq:eq17}
\bar{r}_n(t)= \begin{cases}
\eqref{eq:eq16}, & \text{legal action}, \\
-1, & \text{illegal action}.
\end{cases}
\end{equation} 

In mobility-controlled $Q$-learning, the state includes: the previous AP coverage region ${\rm AP}_{n-1}$, the current AP coverage region ${\rm AP}_{n}$, and the initial velocity $v_n(l_0)$, which is expressed as
\begin{equation}\label{eq:eq18}
s_n=\left\{{\rm AP}_{n-1}, {\rm AP}_{n}, v_n(l_0) \right\}.
\end{equation}
The action is the target velocity, expressed as
\begin{equation}\label{eq:eq19}
a_n=\left\{ v_{{\rm goal},n} \right\}.
\end{equation}
The reward is the accumulated instantaneous reward $\bar{r}_n(t)$ in an AP coverage region, formulated as 
\begin{equation}\label{eq:eq20}
r_n=\sum\limits^{L_n}_{l=1} \bar{r}_n(l).
\end{equation}

The details of the dual-agent $Q$-learning algorithm for joint velocity control and task offloading are shown in Algorithm \ref{alg:alg1}. 
In this algorithm, the information of the robot-located coverage region (sequentially increasing from 1 to $N$) and offloaded data size is leveraged for service migration. 

{\bf Remark:}
The relationship between task offloading and robot mobility is shown by the difference ${k_n} T_{\rm move}-\Delta T {L_n}$.
Since the increased velocity decreases the number of offloading intervals $L_n$, when the reduced offloading time is smaller than the moving time per AP coverage region, that is, $\Delta T L_n < k_n T_{\rm move}$, a positive reward can be obtained.
Otherwise, a negative/zero reward will be incurred.

In the following, the convergence of Algorithm \ref{alg:alg1} will be analyzed.
According to \cite{melo2001convergence}, when $\sum^{T_{\rm Epi}}_{j=1} \lambda_{j}=\infty$ and $\sum^{T_{\rm Epi}}_{j=1} \lambda^{2}_{j}<\infty$, the $Q$ function can converge to the optimal $Q$ function $Q^{*}$ based on the following update rule
\begin{align}\label{EQ:EQ5}
	Q_{j+1}(s_j, a_j) =  & Q_{j}(s_j, a_j) + \lambda_{j} \big( r_{j} + \max\limits_{b \in \mathcal{A}} Q_{j}(s_{j+1}, b) \nonumber \\
	& - Q_{j}(s_j, a_j) \big),
\end{align}
where $\mathcal{A}$ is the set of action spaces.
In our proposed Algorithm \ref{alg:alg1}, two $Q$ functions are updated, where one is for task offloading and the other is for velocity control.
Since the update rule in \eqref{EQ:EQ5} is utilized for two $Q$ functions and $\lambda_{j} = \lambda$ $(0< \lambda <1)$, the optimal $Q$ functions can be obtained separately from these two $Q$ tables. 
Thus, Algorithm \ref{alg:alg1} is convergent.
In the sequel, we specify the convergence of Algorithm \ref{alg:alg1}.

\begin{algorithm}[t] 
	\renewcommand{\algorithmicrequire}{\textbf{Input:}}
	\renewcommand{\algorithmicensure}{\textbf{Output:}}
	\caption{\small Joint Task Offloading and Velocity Control Based on a Dual-Agent $Q$-Learning Algorithm} 
	\label{alg:alg1} 
	\begin{algorithmic}[1]
		\REQUIRE Initialize the table entry $Q_1(s,a)=0$ and $Q_2(s,a)=0$, velocity range $[v_{\rm min}, v_{\rm max}]$, moving distance $c_n$, learning rate $\lambda$, greedy factor $\epsilon$, discount factor $\gamma$.
		\ENSURE Offloading decision $\textbf{1}_{\alpha}(m,t)$, target velocity $v_{{\rm goal},n}$.
		\FOR {$j=1,2,\ldots,T_{\rm Epi}$}
		\STATE Reset $t=1$ and $s_n(t)$;
		\FOR {$n=1,2,\ldots,N$}
		\STATE Observe state $s_n$;
		\STATE Chose action $a_n$ with $\epsilon$-greedy algorithm;
		\WHILE {${\rm AP}(t)=n$}
		\STATE Observe state $s_n(t)$;
		\STATE Chose action $a_n(t)$ with $\epsilon$-greedy algorithm;
		\STATE Calculate reward $\bar{r}_n(t)$ and next state $s_n(t+1)$;
		\STATE Update the $Q$-table for task offloading:
		\begin{align}\label{eq:eq21}
			Q_1(s_n(t),a_n(t)) = & Q_1(s_n(t),a_n(t)) \!+\! \lambda \big(\bar{r}_n(t) \!+\! \gamma  \nonumber \\
			& \cdot \max Q_1(s_n(t\!+\!1),a_n(t\!+\!1))  \nonumber \\
			& - Q_1(s_n(t),a_n(t)) \big)
		\end{align}
		\STATE Update state $s_n(t)=s_n(t+1)$;
		\STATE $t=t+1$;
		\ENDWHILE
		\STATE Calculate reward $r_n$ and next state $s_{n+1}$;
		\STATE Update the $Q$-table for velocity control:
		\begin{align}\label{eq:eq22}
			Q_2(s_n,a_n) = & Q_2(s_n,a_n)+\lambda \big(r_n -Q_2(s_n,a_n)  \nonumber \\
			& + \gamma \max Q_2(s_{n+1},a_{n+1}) \big)
		\end{align}
		\STATE Update state $s_n=s_{n+1}$.
		\ENDFOR
		\ENDFOR		
	\end{algorithmic} 
\end{algorithm}

According to \cite{NEURIPS2020_c20bb2d9}, we have the first convergence theorem as follows. 
\begin{theorem}\label{Theo:Theo3}
Assume that $\| Q_{1,1}( s_{n}(t), a_{n}(t) )\| \leqslant \frac{r_{\rm max}}{1-\gamma}$ and $\| Q_{2,1}( s_{n}, a_{n})\| \leqslant \frac{L_{\rm max} r_{\rm max}}{1-\gamma}$. 
Then, one has 
\begin{equation}\label{EQ:EQ6}
\| Q_{1,j}( s_{n}(t), a_{n}(t) )\| \leqslant \frac{r_{\rm max}}{1-\gamma},
\end{equation}
\begin{equation}\label{EQ:EQ7}
\| Q_{1,j}( s_{n}(t), a_{n}(t) ) - Q^{*}_{1} \| \leqslant \frac{2 r_{\rm max}}{1-\gamma},
\end{equation}
\begin{equation}\label{EQ:EQ8}
\| Q_{2,j}( s_{n}, a_{n})\| \leqslant \frac{L_{\rm max} r_{\rm max}}{1-\gamma},
\end{equation}
\begin{equation}\label{EQ:EQ9}
\| Q_{2,j}( s_{n}, a_{n}) - Q^{*}_{2} \| \leqslant \frac{2 L_{\rm max} r_{\rm max}}{1-\gamma},
\end{equation}
where $r_{\rm max}=\max\limits_{n}\| \bar{r}_n(t) \|$, $L_{\rm max}=\max\limits_{n} L_{n}$, $j=1,2,\ldots,T_{\rm Epi}$, $n=1,2,\ldots,N$, $Q^{*}_{1}$ and $Q^{*}_{2}$ denote the optimal functions of $Q_{1,j}( s_{n}(t), a_{n}(t) )$ and $Q_{2,j}( s_{n}, a_{n})$, respectively. 
\end{theorem}

\begin{IEEEproof}
Mathematical induction is employed to prove Theorem \ref{Theo:Theo3}.
First, for $j=1$, the two initialized $Q$ tables satisfy $\| Q_{1,1}( s_{n}(t), a_{n}(t) )\| \leqslant \frac{r_{\rm max}}{1-\gamma}$ and $\| Q_{2,1}( s_{n}, a_{n})\| \leqslant \frac{L_{\rm max} r_{\rm max}}{1-\gamma}$.
For example, the initial values of two $Q$ tables can be from the intervals $\left[-\frac{r_{\rm max}}{1-\gamma}, \frac{r_{\rm max}}{1-\gamma} \right]$ and $\left[ -\frac{L_{\rm max} r_{\rm max}}{1-\gamma}, \frac{L_{\rm max} r_{\rm max}}{1-\gamma} \right]$.
When $\| Q_{1,j}( s_{n}(t), a_{n}(t) )\| \leqslant \frac{r_{\rm max}}{1-\gamma}$, for the $(j+1)$th iteration, we derive 
\begin{align}\label{EQ:EQ10}
&\| Q_{1,j+1}( s_{n}(t), a_{n}(t) )\|   \nonumber \\
&\quad \leqslant (1-\lambda) \| Q_{1,j}( s_{n}(t), a_{n}(t) ) \| + \lambda \| \bar{r}_{n}(t) \|   \nonumber \\
 &\quad \quad + \lambda \gamma \max\limits_{a_n(t+1) \in \mathcal{A}_1} \| Q_{1,j}( s_{n}(t+1), a_{n}(t+1) ) \| \nonumber \\
&\quad \leqslant \frac{r_{\rm max}}{1-\gamma} + \lambda r_{\rm max} + \lambda \gamma \frac{r_{\rm max}}{1-\gamma}  \nonumber \\
&\quad =\frac{r_{\rm max}}{1-\gamma}.
\end{align}
Thus, \eqref{EQ:EQ6} is proven.

Similarly, when $\| Q_{2,j}( s_{n}, a_{n} )\| \leqslant \frac{L_{\rm max} r_{\rm max}}{1-\gamma}$, for the $(j+1)$th iteration, we obtain 
 \begin{align}\label{EQ:EQ11}
\| Q_{2,j+1}( s_{n}, a_{n} )\| 
& \leqslant (1-\lambda) \| Q_{2,j}( s_{n}, a_{n} ) \| + \lambda \| r_{n} \|  \nonumber \\
 & \quad + \lambda \gamma \max\limits_{a_{n+1} \in \mathcal{A}_2} \| Q_{2,j}( s_{n+1}, a_{n+1} ) \| \nonumber \\
& \leqslant \frac{L_{\rm max} r_{\rm max}}{1-\gamma} \!+\! \lambda L_{\rm max} r_{\rm max} \!+\! \lambda \gamma \frac{L_{\rm max} r_{\rm max}}{1-\gamma}  \nonumber \\
& =\frac{L_{\rm max} r_{\rm max}}{1-\gamma}.
\end{align}
Thus, \eqref{EQ:EQ8} is proven.

Based on the above proof, \eqref{EQ:EQ7} and \eqref{EQ:EQ9} can be proven as 
\begin{align}\label{EQ:EQ12}
\| Q_{1,j}( s_{n}(t), a_{n}(t) ) - Q^{*}_{1} \| & \leqslant \| Q_{1,j}( s_{n}(t), a_{n}(t) ) \| + \| Q^{*}_{1} \|   \nonumber \\
 & \leqslant  \frac{2 r_{\rm max}}{1-\gamma},
\end{align}
\begin{equation}\label{EQ:EQ13}
\| Q_{2,j}( s_{n}, a_{n}) - Q^{*}_{2} \| \leqslant \| Q_{2,j}( s_{n}, a_{n}) \| + \| Q^{*}_{2} \|  \leqslant \frac{2 L_{\rm max} r_{\rm max}}{1-\gamma}.
\end{equation}

This completes the proof.
\end{IEEEproof}

Theorem \ref{Theo:Theo3} shows that, with the reduced value of $\gamma$, the convergence performance of Algorithm \ref{alg:alg1} will be improved. 
Based on \cite{pmlr-v120-lee20a}, Theorem \ref{Theo:Theo3} can be extended to the second convergence theorem.
First, the Bellman operation $\mathcal{T}\left\{ \cdot \right\}$ is defined as
\begin{equation}\label{EQ:EQ14}
\mathcal{T}\left\{ Q(s,a) \right\} \!=\! \sum\limits_{s^{\prime} \in \mathcal{S}} p_{a}(s,s^{\prime}) \Big( \!r(s,a) + \gamma \max\limits_{a^{\prime} \in \mathcal{A}} Q(s^{\prime},a^{\prime}) \! \Big),
\end{equation}
where $p_{a}(s,s^{\prime})$ is the state transition probability from State $s$ to State $s^{\prime}$ and $\mathcal{S}$ is the set of state spaces.
Then, an approximation error $\theta_{i,j}$ $(i=1,2, j \in [1, T_{\rm Epi}])$ is defined as
\begin{equation}\label{EQ:EQ15}
E\left\{ \left\| Q_{i,j+1} - \mathcal{T}\left\{ Q_{i,j} \right\} \right\|^2_2 \right\} \leqslant \theta_{i,j}.
\end{equation}
Based on Theorem \ref{Theo:Theo3}, we assume that $\theta_{2,j}=L_{\rm max}\theta_{1,j}$.

\begin{theorem}\label{Theo:Theo4}
The convergence of the improved $Q$-learning in Algorithm \ref{alg:alg1} can be expressed as
\begin{align}\label{EQ:EQ16}
& E\left\{ \left\| Q_{1,T_{\rm Epi}} (s_n(t), a_n(t)) - Q^{*}_{1} \right\|_{\infty} \right\}  \nonumber \\
& \leqslant   \sum\limits^{T_{\rm Epi}}_{j=1} \! \gamma^{T_{\rm Epi} \!-\! j} \sqrt{\theta_{1,j}}  
\!+\! \gamma^{T_{\rm Epi}} E\!\left\{\! \left\| Q_{1,0} (s_n(t), a_n(t)) \!-\! Q^{*}_{1} \right\|_{\infty} \!\right\}\!,
\end{align}
\begin{align}\label{EQ:EQ17}
& E\left\{ \left\| Q_{2,T_{\rm Epi}} (s_n, a_n) - Q^{*}_{2} \right\|_{\infty} \right\}  \nonumber \\
& \leqslant  \sum\limits^{T_{\rm Epi}}_{j=1} \gamma^{T_{\rm Epi} \!-\! j} \sqrt{L_{\rm max} \theta_{1,j}}  
\!+\! \gamma^{T_{\rm Epi}} E\!\left\{ \left\| Q_{2,0} (s_n, a_n) \!-\! Q^{*}_{2} \right\|_{\infty} \right\}\!.
\end{align}
When $\theta_{1,j}=\theta$, \eqref{EQ:EQ16} and \eqref{EQ:EQ17} can be rewritten by
\begin{align}\label{EQ:EQ18}
& E\left\{ \left\| Q_{1,T_{\rm Epi}} (s_n(t), a_n(t)) - Q^{*}_{1} \right\|_{\infty} \right\}  \nonumber \\
& \leqslant  \frac{\sqrt{\theta}}{1-\gamma} 
+ \gamma^{T_{\rm Epi}} E\left\{ \left\| Q_{1,0} (s_n(t), a_n(t)) - Q^{*}_{1} \right\|_{\infty} \right\},
\end{align}
\begin{align}\label{EQ:EQ19}
& E\left\{ \left\| Q_{2,T_{\rm Epi}} (s_n, a_n) - Q^{*}_{2} \right\|_{\infty} \right\}  \nonumber \\
& \leqslant  \frac{\sqrt{L_{\rm max} \theta}}{1-\gamma} 
+ \gamma^{T_{\rm Epi}} E\left\{ \left\| Q_{2,0} (s_n, a_n) - Q^{*}_{2} \right\|_{\infty} \right\}.
\end{align}
\end{theorem}

\begin{IEEEproof}
Considering the $\gamma$-contraction property of the Bellman operator and $Q^{*}_{i}=\mathcal{T}\{ Q^{*}_{i} \}$ in \cite{NEURIPS2020_c20bb2d9}, we can derive
\begin{align}\label{EQ:EQ20}
& E\left\{ \left\| Q_{i,j+1}  - Q^{*}_{i} \right\|_{\infty} \right\}  \nonumber \\
& \leqslant  E\left\{ \left\| Q_{i,j+1}  - \mathcal{T} \left\{ Q_{i,j} \right\} \right\|_{\infty} \right\}  +  E\left\{ \left\|  \mathcal{T} \left\{ Q_{i,j} \right\} - Q^{*}_{i} \right\|_{\infty} \right\}  \nonumber \\
& \leqslant \sqrt{E\!\left\{ \left\| Q_{i,j\!+\!1}  \!-\! \mathcal{T} \! \left\{ Q_{i,j} \right\} \right\|^2_2 \right\}} \!+\!  E\!\left\{ \left\|  \mathcal{T} \left\{ Q_{i,j} \right\} \!-\! \mathcal{T} \left\{ Q^{*}_{i} \right\} \right\|_{\infty} \right\}  \nonumber \\
& \leqslant \sqrt{\theta_{i,j+1}} + \gamma  E\left\{ \left\|   Q_{i,j}  - Q^{*}_{i} \right\|_{\infty} \right\}.
\end{align}
Based on \eqref{EQ:EQ20}, we can derive
\begin{align}\label{EQ:EQ21}
& E\left\{ \left\| Q_{i,T_{\rm Epi}}  - Q^{*}_{i} \right\|_{\infty} \right\}   \nonumber \\
& \leqslant \sqrt{\theta_{i,T_{\rm Epi}}} + \gamma E\left\{ \left\| Q_{i,T_{\rm Epi}-1}  - Q^{*}_{i} \right\|_{\infty} \right\}  \nonumber \\
& \leqslant \sqrt{\theta_{i,T_{\rm Epi}}} \!+\! \gamma \left( \sqrt{\theta_{i,T_{\rm Epi}-1}} \!+\! \gamma E\left\{ \left\| Q_{i,T_{\rm Epi}-2}  \!-\! Q^{*}_{i} \right\|_{\infty} \right\}  \right) \nonumber \\
& \cdots  \nonumber \\
&\leqslant \sum\limits^{T_{\rm Epi}}_{j=1} \gamma^{T_{\rm Epi}-j} \sqrt{\theta_{i,j}} + \gamma^{T_{\rm Epi}}  E\left\{ \left\| Q_{i,0}  - Q^{*}_{i} \right\|_{\infty} \right\}.
\end{align}
Since $\theta_{2,j} = L_{\rm max} \theta_{1,j}$, based on \eqref{EQ:EQ21}, \eqref{EQ:EQ16} and \eqref{EQ:EQ17} can be obtained.

When $\theta_{i,j}=\theta$, \eqref{EQ:EQ21} is simplified as
\begin{align}\label{EQ:EQ22}
& E\left\{ \left\| Q_{i,T_{\rm Epi}}  - Q^{*}_{i} \right\|_{\infty} \right\}    \nonumber \\
& \leqslant \frac{1-\gamma^{T_{\rm Epi}}}{1-\gamma} \sqrt{\theta}  + \gamma^{T_{\rm Epi}}  E\left\{ \left\| Q_{i,0}  - Q^{*}_{i} \right\|_{\infty} \right\}.
\end{align}
Since $0 \leqslant \gamma < 1$, when $T_{\rm Epi} \rightarrow \infty$, we have $\gamma^{T_{\rm Epi}} \rightarrow \infty$. 
Thus, \eqref{EQ:EQ22} is reduced to
\begin{align}\label{EQ:EQ23}
\! E\!\left\{ \! \left\| Q_{i,T_{\rm Epi}}  \!-\! Q^{*}_{i} \right\|_{\infty} \! \right\}    
\! \leqslant \! \frac{\sqrt{\theta}}{1 \!-\! \gamma} \!+\! \gamma^{T_{\rm Epi}}  E\!\left\{\! \left\| Q_{i,0}  \!-\! Q^{*}_{i} \right\|_{\infty} \! \right\}\!.
\end{align}
Finally, according to \eqref{EQ:EQ23}, \eqref{EQ:EQ18} and \eqref{EQ:EQ19} can be proved.
\end{IEEEproof}

Theorem \ref{Theo:Theo4} shows that the convergence of the proposed algorithm is affected by four factors: 1) the approximation error for $Q_1$; 2) the approximation calculation of the Bellman operator for $Q_1$; 3) the approximation error for $Q_2$; and 4) the approximation calculation of the Bellman operator for $Q_2$. 

In the following, the complexity of the proposed dual-agent $Q$-learning is analyzed in terms of sample complexity.
The sample complexity is referred to as the total number of samples needed to yield an entrywise $\xi_1$-accurate approximation of the optimal $Q$-function, to satisfy $\max_{s,a} \| Q(s,a) - Q^{*}(s,a) \| \leqslant \xi_1$ (or $E\{\max_{s,a} \| Q(s,a) - Q^{*}(s,a) \|\} \leqslant \xi_1$) with probability at least $1-\xi_2$ for any $\xi_2 \in (0,1)$ \cite{li2021q}. 
According to \cite{li2021q}, we assume a $\gamma$-discounted infinite-horizon MDP with state space $\mathcal{S}$ and action space $\mathcal{A}$.
We also assume that the Markov chain induced by a behavior policy $\pi_b$ is uniformly ergodic.
The minimum state-action occupancy probability of the sample trajectory is defined as $\omega_{\min,i}=\min\limits_{(s,a)\in \mathcal{S}_i\times \mathcal{A}_i} \omega_{\pi_b,i}(s,a)$, where $\omega_{\pi_b,i}$ denotes the stationary distribution of the Markov chain for the $i$th agent with $i=1,2$.
Furthermore, the mixing time associated with the sample trajectory is defined as $t_{{\rm mix},i}=\min\left\{ t \left| \max\limits_{(s,a)\in \mathcal{S}_i \times \mathcal{A}_i} d_{\rm TV}\left(P^{t}(\cdot | s,a), \omega_{\pi_b,i} \right) \leq \frac{1}{4} \right.\right\}$, where $d_{\rm TV}(\omega,\nu) = 0.5 \sum_{x \in \mathcal{X}} |\omega(x) - \nu(x)|$ and $P^{t}(\cdot | s,a)$ indicates the distribution of $(s_t, a_t)$ with the initialization of $(s_0, a_0) = (s, a)$.
According to Theorem \ref{Theo:Theo3}, in dual-agent $Q$-learning, the accuracy levels of the two agents are $\frac{2 r_{\rm max}}{1-\gamma}$ and $\frac{2 L_{\rm max} r_{\rm max}}{1-\gamma}$.
According to the complexity analysis of asynchronous $Q$-learning in \cite{li2021q}, the total sample complexity of dual-agent $Q$-learning can be approximated as the sum of the sample complexity of the offloading agent and that of the velocity control agent as
\begin{align}\label{EQ:EQ24}
O\bigg( & \frac{1}{4 r^2_{\rm max} (1-\gamma)^2} \left(\frac{1}{\omega_{\min,1}} + \frac{1}{L^2_{\rm max} \omega_{\min,2}}\right)  \nonumber \\
   	&	+ \frac{1}{1-\gamma} \left(\frac{t_{\rm mix,1}}{\omega_{\min,1}} + \frac{t_{\rm mix,2}}{\omega_{\min,2}} \right)  \bigg).
\end{align}

\section{Simulation Results and Discussion}

In this section, we will evaluate the stochastic task offloading performance achieved from our proposed $Q$-learning-based algorithm.

In our simulation, 20 APs and MEC servers are deployed, where $\mathcal{N}_1=\left\{ 1,2,\ldots,7 \right\} \cup \left\{ 14,15,\ldots,20\right\}$ and $\mathcal{N}_2=\left\{ 8,9,\ldots,13 \right\}$.
The computing capacities of cellular and satellite-based MEC servers are from finite sets $\{10,11,\ldots,19 \}$ (GHz) and $\{50,51,\ldots,59 \}$ (GHz).
The moving distance $c_n$ is randomly chosen from a set $\{100,200,300\}$ (m) for the cellular AP and from a set $\{1000,2000,3000\}$ (m) for the satellite.
In the cellular network, the bandwidth is $W=10$ MHz, the transmit power of the mobile robot is $p=0.2$ W, the channel noise power is $\sigma^2=2\times 10^{-12}$ W, and the channel power gain is $h^2=10^{-6}$.
In the satellite communication, for simplicity, we set the distances as $d_{\rm GS}=d_{\rm SE}=1000$ km, and the transmission rates as $r_{\rm GS}=10$ Mbps and $r_{\rm SE}=100$ Mbps.
The extra migration cost $\Delta C$ is set to the average delay of 500 ms.
In each offloading interval $\Delta T=1$ s, the generated data size is randomly selected from the set $\{100,250,400,550,700\}$ (KB) with $\Phi=800$ CPU cycles/bit, and the computing capacity of the mobile robot is randomly chosen from a finite set $\{0.5,0.6,\ldots,1\}$ (GHz).
The random waypoint model without considering direction \cite{5949158} is used as the mobility model of the robot.
For the mobile robot with $a=2 \; {\rm m/s}^2$, its velocity is from a discrete set $\mathcal{V}=\{5,6,\ldots,20\}$ (m/s).
In $Q$-learning, the hyperparameters are set as $\lambda=0.1$, $\gamma=0.9$, and $\epsilon=0.05$ with a discount interval of $4\times 10^{-6}$.

For a fair performance comparison, we simulate three baselines below.
\begin{itemize}
  \item {\it Conventional Offloading:} 
  The mobile robot has a constant velocity, as depicted in \cite{9685350}, and its offloading decision is made by $Q$-learning.
  \item {\it Local Execution:} 
  All scheduled computation tasks are processed by a mobile robot with its available CPU frequency while maintaining a constant velocity. 
  \item {\it Simplified Greedy:} 
  Since conventional greedy searching for velocity decision-making has significant complexity, i.e., $O(|\mathcal{V}|^{N})$, a simplified greedy algorithm using local searching for each AP is given for comparison.
  First, in the $n$th AP coverage region, the velocity with the maximal average of $\bar{r}_n(t)$ is searched from the candidate set $\mathcal{V}$.
  Then, upon searching the maximal average of $r_n$ in all training, the target velocity for the whole trajectory will be selected.  
\end{itemize}

\begin{figure}[t]
	\setlength{\abovecaptionskip}{0.cm}
	\setlength{\belowcaptionskip}{-0.cm}
	\centerline{\includegraphics[width=3.1in]{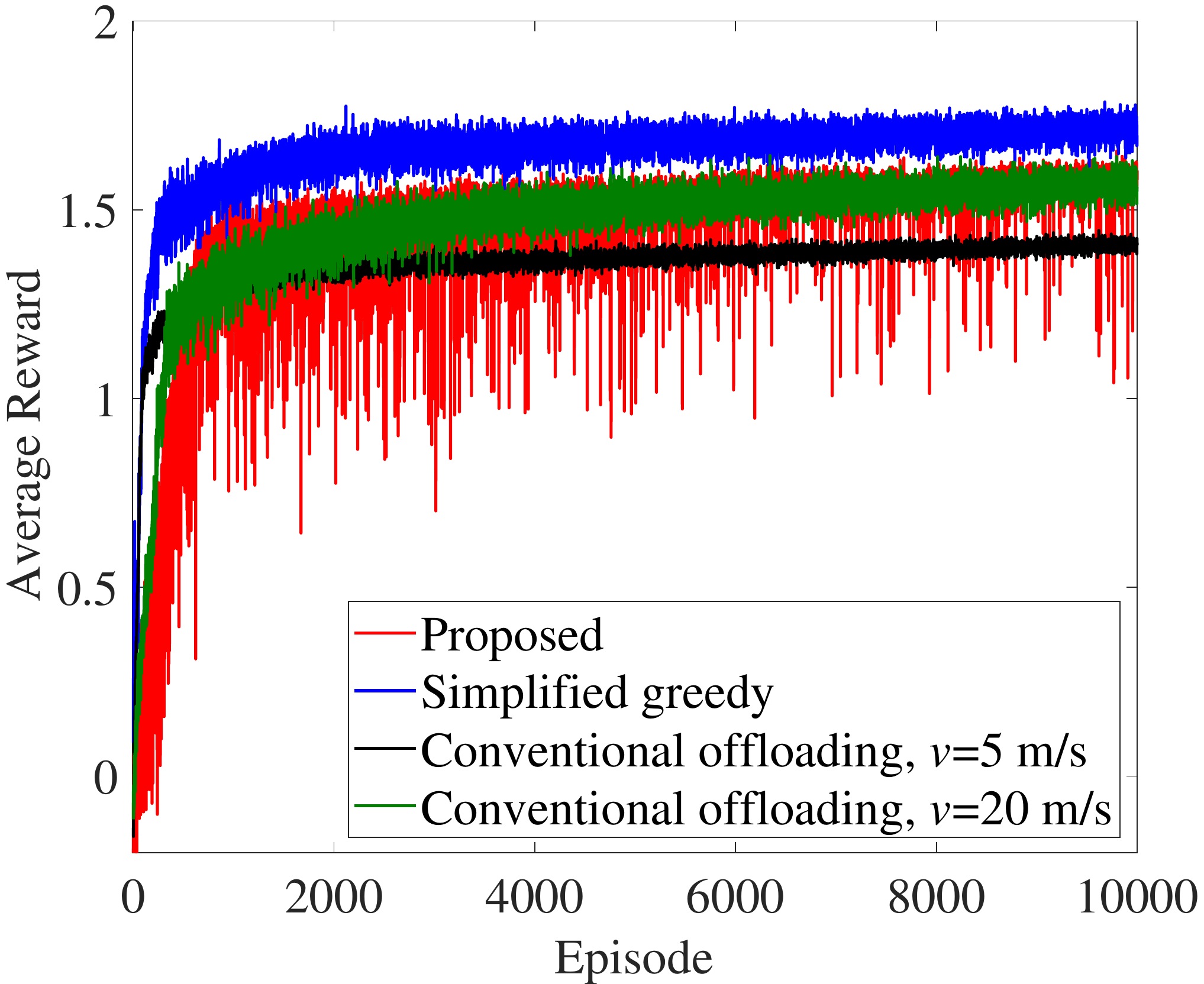}}
	\caption{Average reward in the training episode. 
		The number of APs with unavailable wireless communication is $N_{\rm CH}=4$. 
		The migration ratio is $\rho=0.1$.
		The moving factor is $\theta=0.1$.}
	\label{Fig:Fig3}
\end{figure}

\begin{figure}[!tp]
	\centering
	\subfloat[]{\includegraphics[width=3.2in]{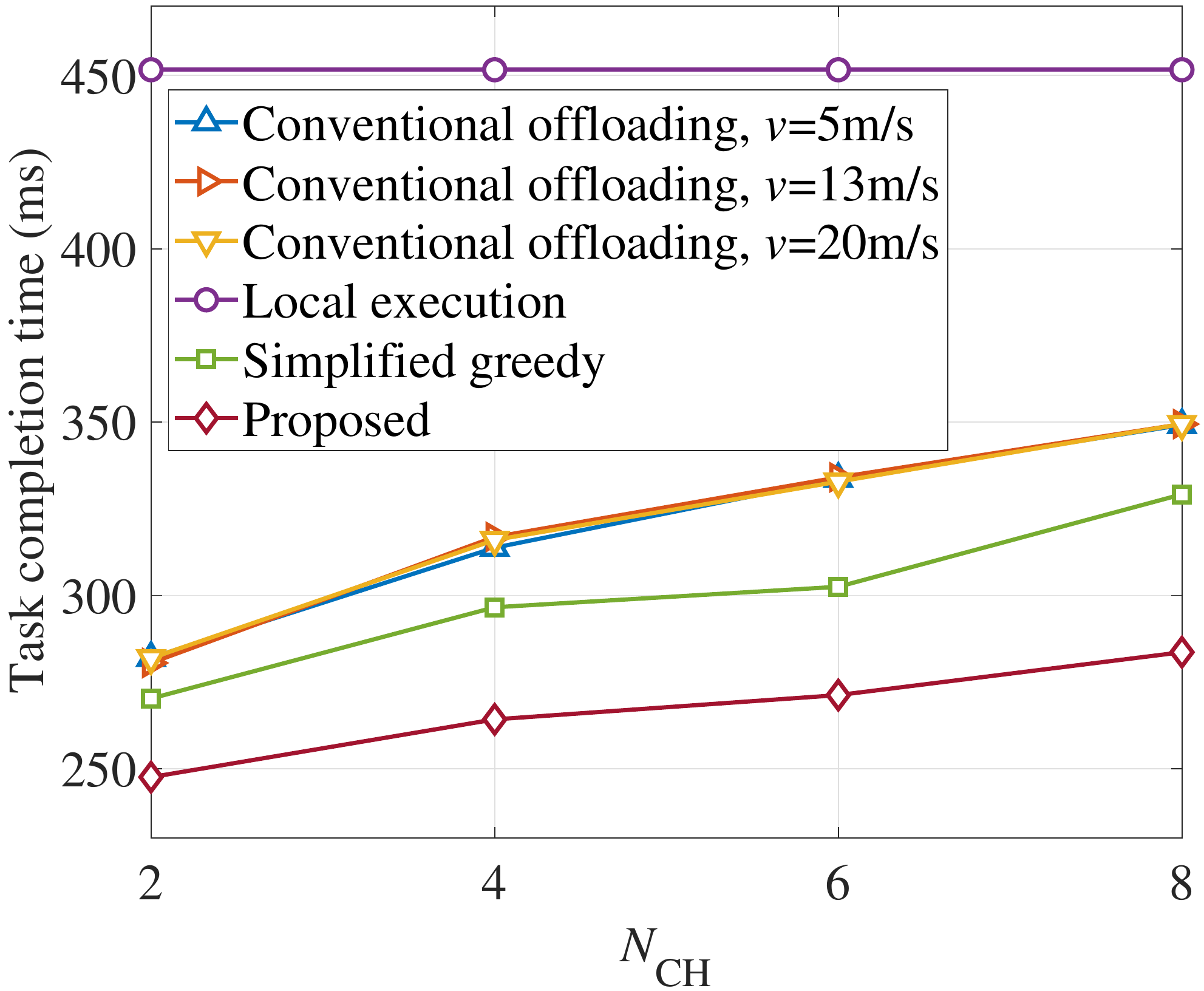}
		\label{Fig:Fig4a}}
	\hfil
	\subfloat[]{\includegraphics[width=3.2in]{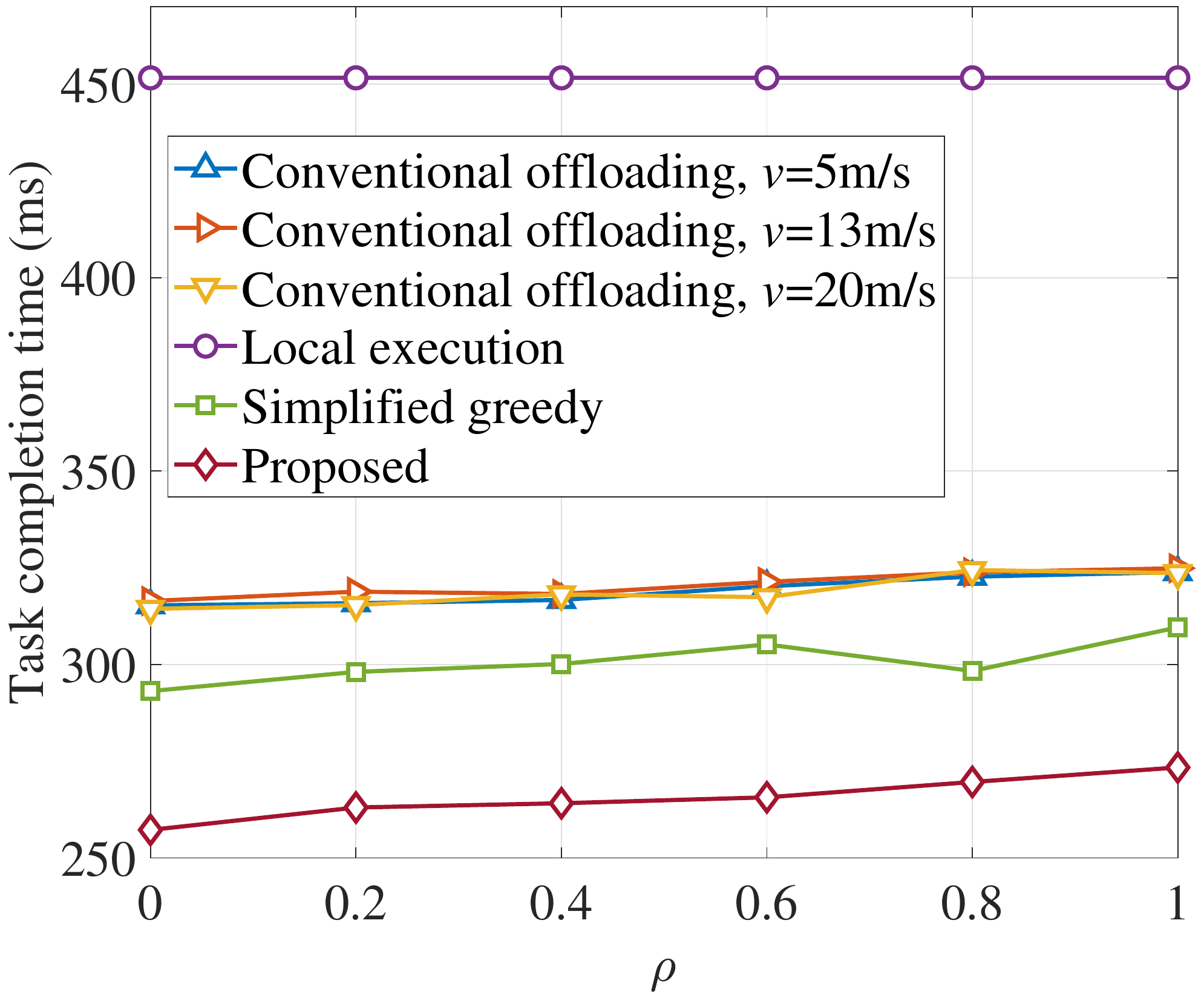}
		\label{Fig:Fig4b}}
	\hfil
	\subfloat[]{\includegraphics[width=3.2in]{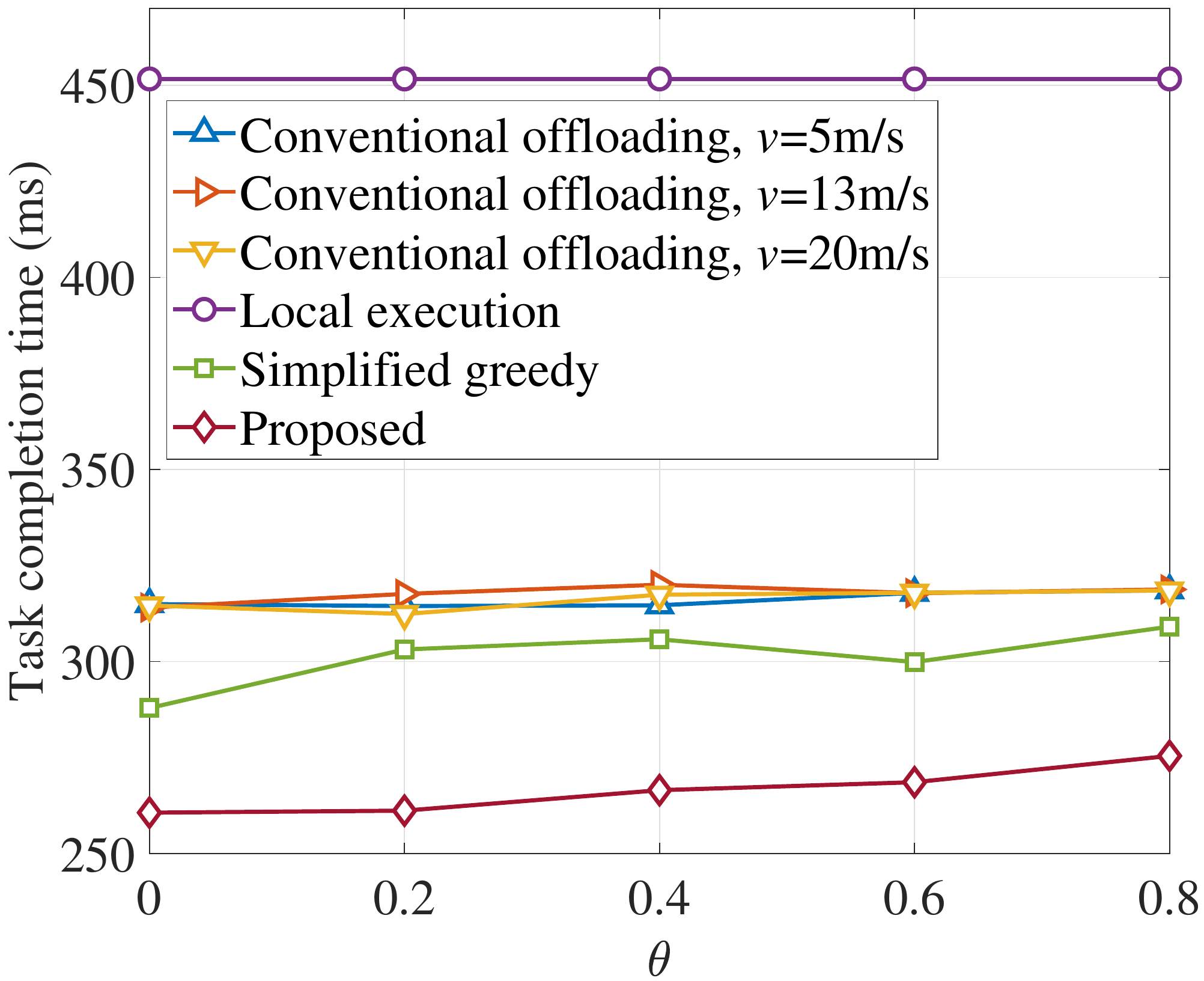}
		\label{Fig:Fig4c}}
	\caption{Task completion time comparison among conventional offloading, local execution, simplified greedy, and the proposed schemes. (a) $\rho=0.1$ and $\theta=0.1$, (b) $N_{\rm CH}=4$ and $\theta=0.1$, (c) $N_{\rm CH}=4$ and $\rho=0.1$.}
	\label{Fig:Fig4}
\end{figure}

In Fig. \ref{Fig:Fig3}, the convergence of the reward function in \eqref{eq:eq20} is plotted for the proposed scheme compared to conventional offloading and simplified greedy schemes.
The number of cellular/satellite APs with unavailable wireless communication is set to $N_{\rm CH}=4$. 
The migration ratio is set as $\rho=0.1$.
The moving factor is set as $\theta=0.1$.
As observed in Fig. \ref{Fig:Fig3}, the proposed scheme has higher rewards than conventional offloading using low velocities and has close rewards to conventional offloading using the highest velocity.
It also has slightly lower rewards than the simplified greedy algorithm.
Thus, the proposed scheme can achieve convergence as the training increases, which is consistent with the convergence analysis in Section III-B.

In Fig. \ref{Fig:Fig4}, the average task completion time of conventional offloading, local execution, simplified greedy, and the proposed schemes are plotted for different parameters.
We first compare the proposed scheme with conventional schemes in Fig. \ref{Fig:Fig4}\subref{Fig:Fig4a} for different values of $N_{\rm CH}$.
It can be seen in Fig. \ref{Fig:Fig4}\subref{Fig:Fig4a} that as the value of $N_{\rm CH}$ increases, except for the local execution, the other schemes have increased task completion times.
More importantly, the proposed scheme has a lower completion time than conventional schemes.
For example, for $N_{\rm CH}=4$, compared to conventional offloading, local execution, and simplified greedy, the completion time of the proposed scheme is reduced by approximately 16\%, 41\%, and 11\%, respectively.
We also plot the task completion time curves for varying migration factors $\rho$ in Fig. \ref{Fig:Fig4}\subref{Fig:Fig4b}. 
It is shown that upon increasing the migration ratio, a slightly increased task completion time is incurred.
For varying values of $\rho$, the proposed scheme also has a much lower completion time than conventional offloading, local execution, and simplified greedy, with average reduction ratios of 17\%, 41\%, and 12\%, respectively.
Moreover, in Fig. \ref{Fig:Fig4}\subref{Fig:Fig4c}, we plot the task completion time curves of these schemes for different moving factors $\theta$. 
As seen from this figure, the moving factor has a slight effect on the task completion time of all MEC-based offloading schemes.
In these schemes, the proposed scheme has the lowest task completion time, whose average reduction ratios over conventional offloading, local execution, and simplified greedy are 16\%, 41\%, and 11\%, respectively.
In addition, Fig. \ref{Fig:Fig4} shows that, since velocity control is not considered in the conventional offloading scheme, the conventional scheme has almost the same average task completion time for different velocities.	
To summarize, as opposed to conventional schemes, the proposed scheme can effectively reduce the service delay of MEC-based offloading for the hybrid satellite-terrestrial-network-enabled robot. 

\begin{figure}[t]
	\setlength{\abovecaptionskip}{0.cm}
	\setlength{\belowcaptionskip}{-0.cm}
	\centerline{\includegraphics[width=3.2in]{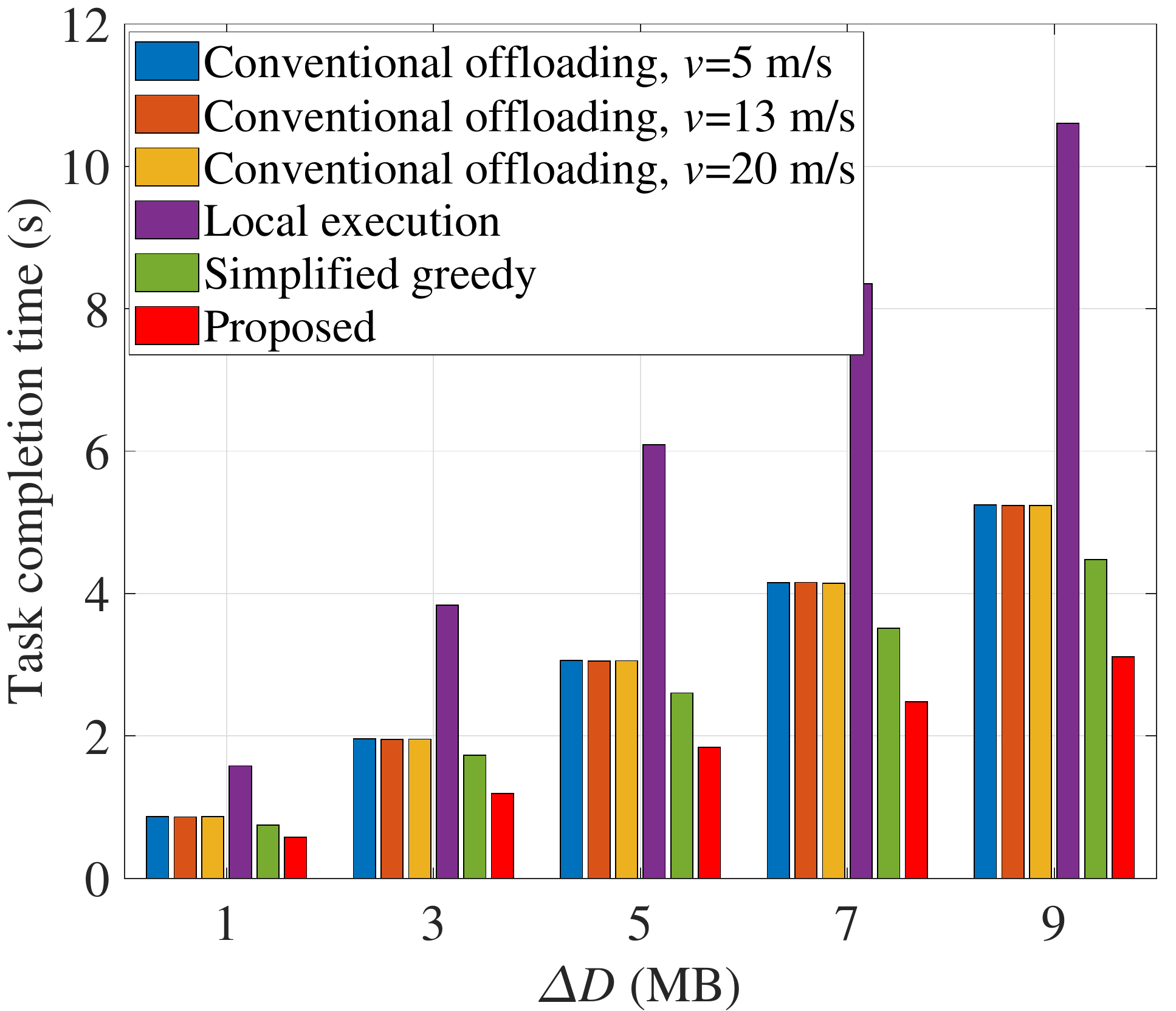}}
	\caption{Task completion time comparison among conventional offloading, local execution, simplified greedy, and the proposed schemes for different data sizes, where $N_{\rm CH}=4$, $\rho=0.1$, and $\theta=0.1$.}
	\label{Fig:Fig5}
\end{figure}

Fig. \ref{Fig:Fig5} portrays the task completion time versus the size of the data generated by the robot.
The data size per offloading interval is randomly selected from the set $\{100,250,400,550,700\}({\rm KB}) + \Delta D$, where the incremental parameter $\Delta D$ belongs to $\{ 1,3,5,7,9\}$ (MB).
It is shown that for different data sizes, the proposed scheme has the lowest task completion time.
Furthermore, compared to conventional schemes, when the data size increases, a much higher time consumption reduction can be achieved by the proposed scheme.

\begin{figure}[tp]
	\centering
	\subfloat[]{\includegraphics[width=3.2in]{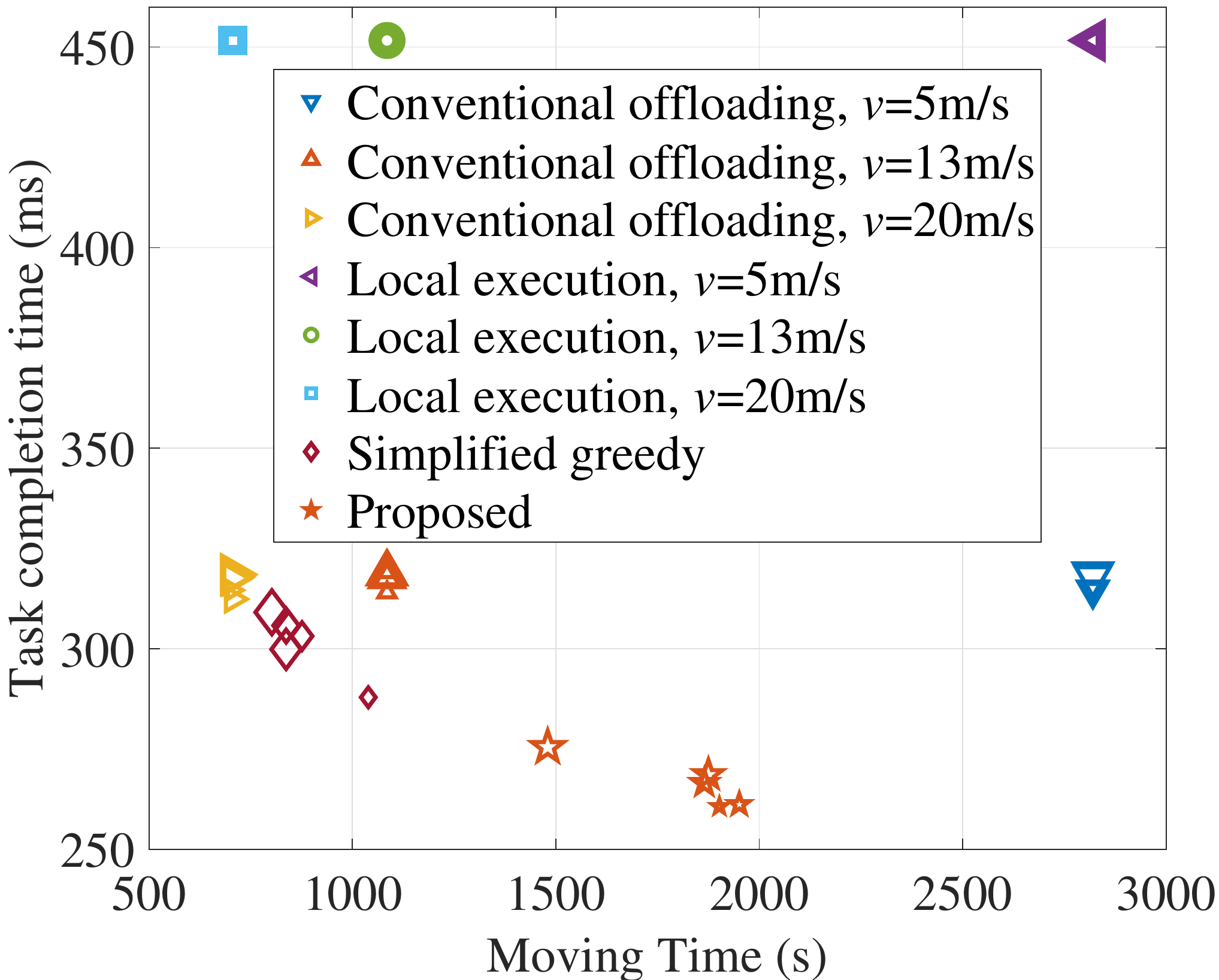}
		\label{Fig:Fig6a}}
	\hfil
	\subfloat[]{\includegraphics[width=3.2in]{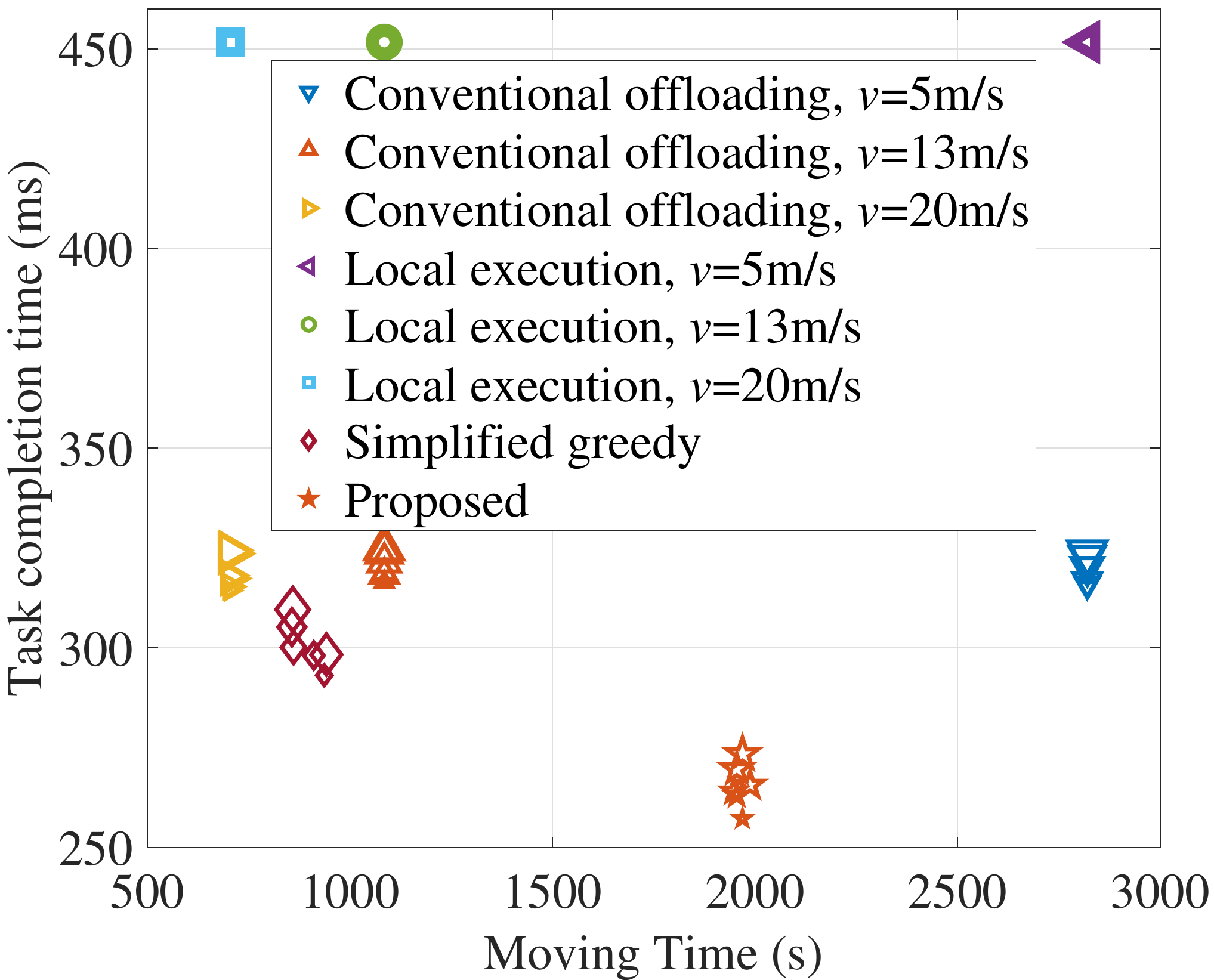}
		\label{Fig:Fig6b}}
	\hfil
	\subfloat[]{\includegraphics[width=3.2in]{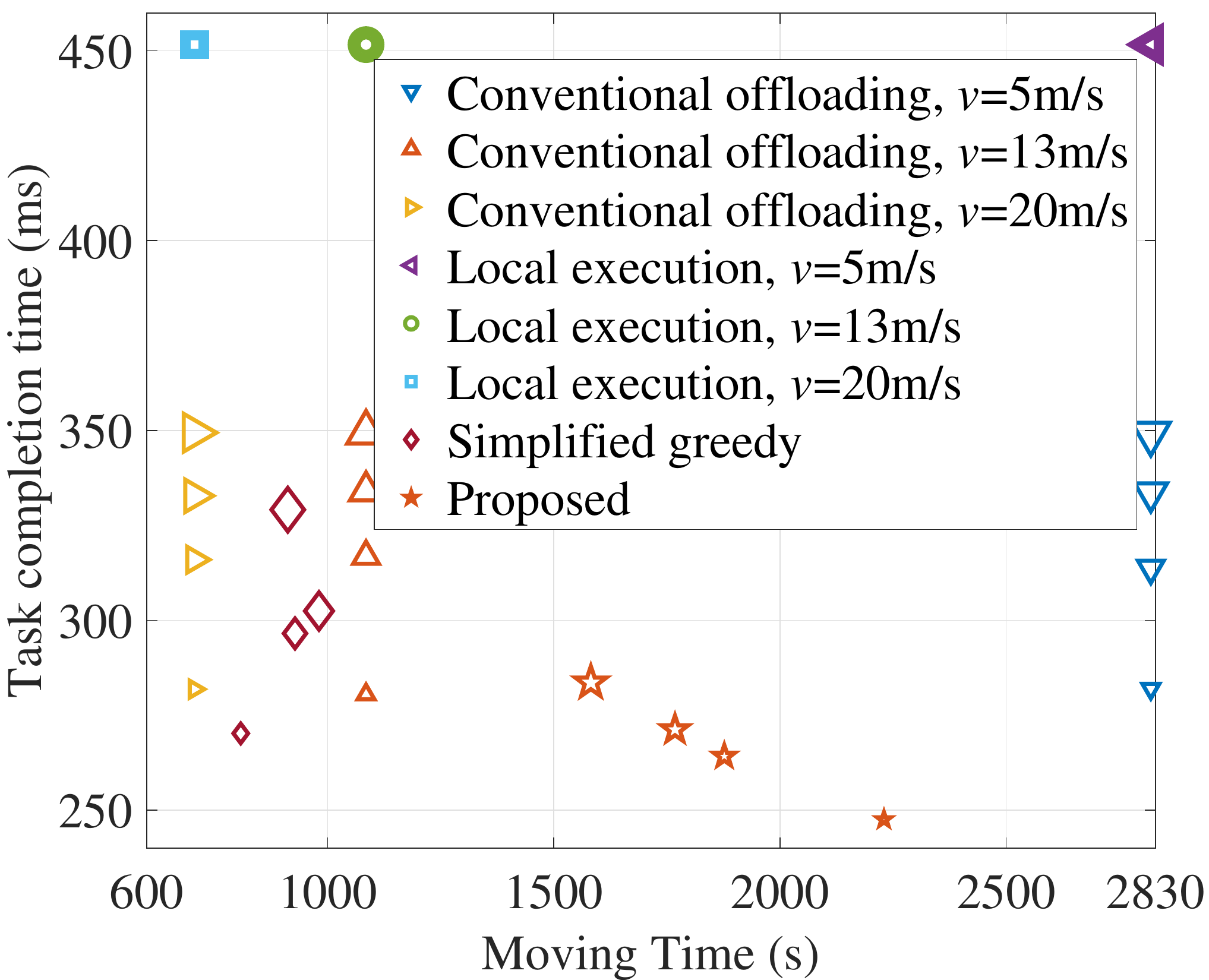}
		\label{Fig:Fig6c}}
	\caption{Task completion time versus moving time among conventional offloading, local execution, simplified greedy, and the proposed schemes. 
		(a) $N_{\rm CH}=2,4,6,8$, $\rho=0.1$, and $\theta=0.1$. When the marker size increases, the value of $N_{\rm CH}$ increases from 2 to 8. 
		(b) $N_{\rm CH}=4$, $\rho=0,0.2,0.4,0.6,0.8,1$, and $\theta=0.1$. When the marker size increases, the value of $\rho$ increases from 0 to 1. 
		(c) $N_{\rm CH}=4$, $\rho=0.1$, and $\theta=0,0.2,0.4,0.6,0.8$. When the marker size increases, the value of $\theta$ increases from 0 to 0.8.}
	\label{Fig:Fig6}
\end{figure}

To show the effect of mobility control on service delay, we now investigate task completion time versus moving time in Fig. \ref{Fig:Fig6}.
It is noted that, when the marker size increases, the values of the parameters involving $N_{\rm CH}$, $\rho$, and $\theta$ are increased.
First, in Fig. \ref{Fig:Fig6}\subref{Fig:Fig6a}, we compare the task completion time versus moving time performance for conventional offloading, local execution, simplified greedy, and the proposed schemes with different values of $N_{\rm CH}$.
It is shown that since a constant velocity is assumed in conventional offloading and local execution, the effect of the velocity control on the completion time performance cannot be clearly observed.
Although the simplified greedy scheme has a lower moving time than the proposed scheme, it has a higher time consumption than the proposed scheme.
Based on the joint optimization of task offloading and velocity control, the proposed scheme obtained the lowest task completion time at a moderate moving time.
Fig. \ref{Fig:Fig6}\subref{Fig:Fig6a} also implies that the proposed scheme is sensitive to $N_{\rm CH}$, since communication state-based velocity control is employed.
Second, we portray a scatterplot to compare the task completion time versus moving time performance for different migration ratios in Fig. \ref{Fig:Fig6}\subref{Fig:Fig6b}.
Similar to \ref{Fig:Fig6}\subref{Fig:Fig6a}, Fig. \ref{Fig:Fig6}\subref{Fig:Fig6b} shows that, compared to conventional schemes, the proposed scheme can achieve the lowest time consumption at a moderate moving time.
It is also shown that compared to the communication state, the migration ratio has a smaller effect on the moving time (corresponding to velocity control) in the proposed scheme.
Finally, in Fig. \ref{Fig:Fig6}\subref{Fig:Fig6c}, we plot the task completion time scatters versus moving time for different moving factors.
As shown in Fig. \ref{Fig:Fig6}\subref{Fig:Fig6c}, the proposed scheme can also achieve the lowest time consumption at a moderate moving time.
Moreover, with an increased moving factor, the proposed scheme can efficiently reduce the moving time while maintaining the lowest service delay among these schemes.
It is also shown in Fig. \ref{Fig:Fig6} that all conventional schemes are insensitive to the moving time, and the proposed scheme benefits from velocity control. 

Furthermore, to verify the efficiency of the proposed scheme, two possible cases for the joint optimization of velocity control and task offloading are considered.
In Case I, according to the availability of all wireless channels in each AP, the mobile robot directly makes a velocity control decision. 
For the channel states $\mu_n=1$ and $\mu_n=0$, we have $v_{{\rm goal},n}=v_{\rm min}$ and $v_{{\rm goal},n}=v_{\rm max}$, respectively. 
In Case II, based on the availability of all wireless channels per AP, the mobile robot directly makes a task offloading decision. 
For the channel states $\mu_n=1$ and $\mu_n=0$, we have $\textbf{1}_{\alpha}(m,t)=1$ with a given MEC server and $\textbf{1}_{\alpha}(m,t)=0$ with the local computation, respectively.
Our proposed scheme corresponds to Case III.
Different from the fixed velocity control in Case I and the fixed offloading decision in Case II, a flexible joint optimization is designed in the proposed scheme.
Thus, Cases I and II can also be regarded as two special cases of Case III.
In the following figures, the task completion time performance is comprehensively compared for these three cases.

\begin{figure}[tp]
	\centering
	\subfloat[]{\includegraphics[width=3.2in]{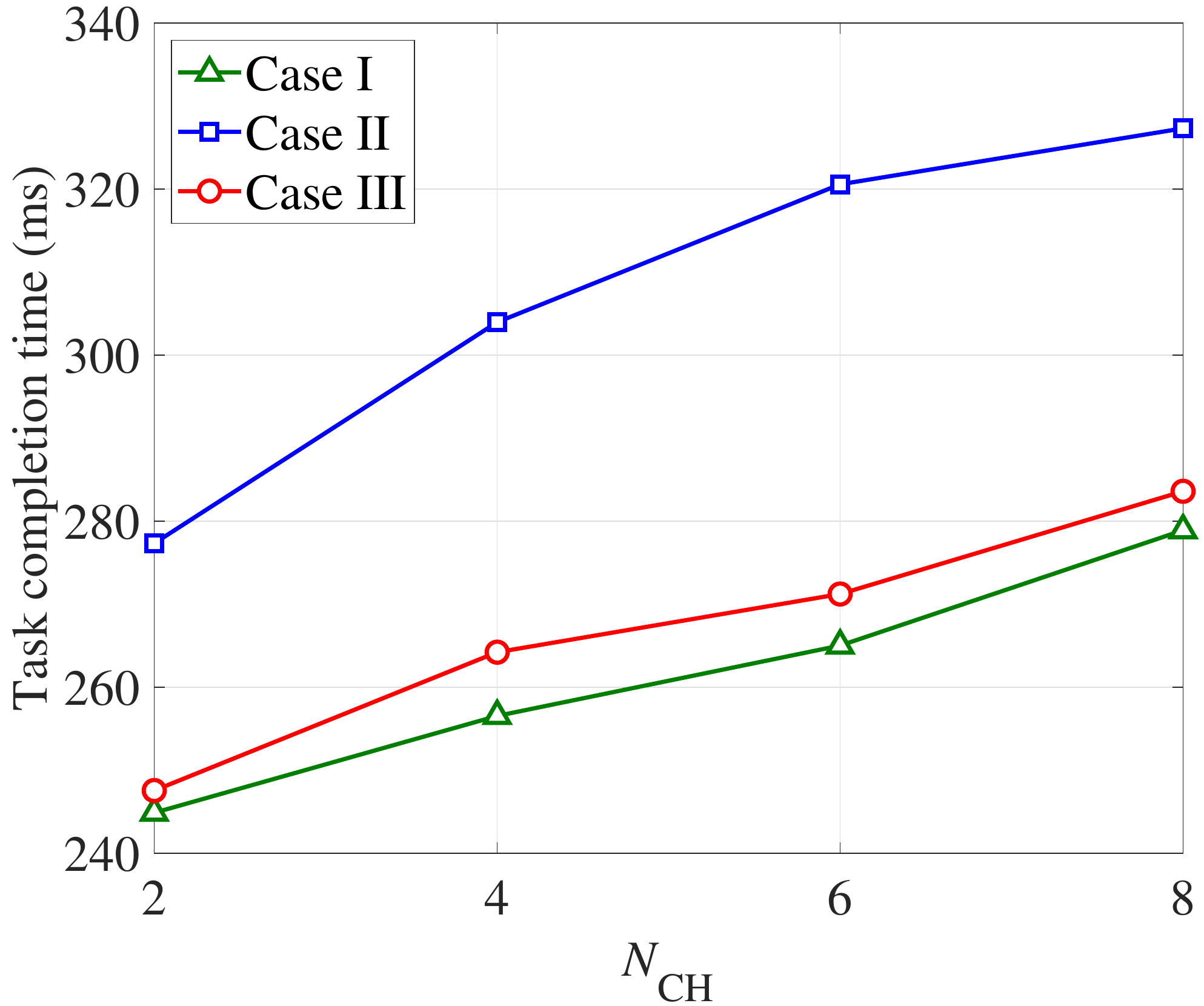}
		\label{Fig:Fig7a}}
	\hfil
	\subfloat[]{\includegraphics[width=3.2in]{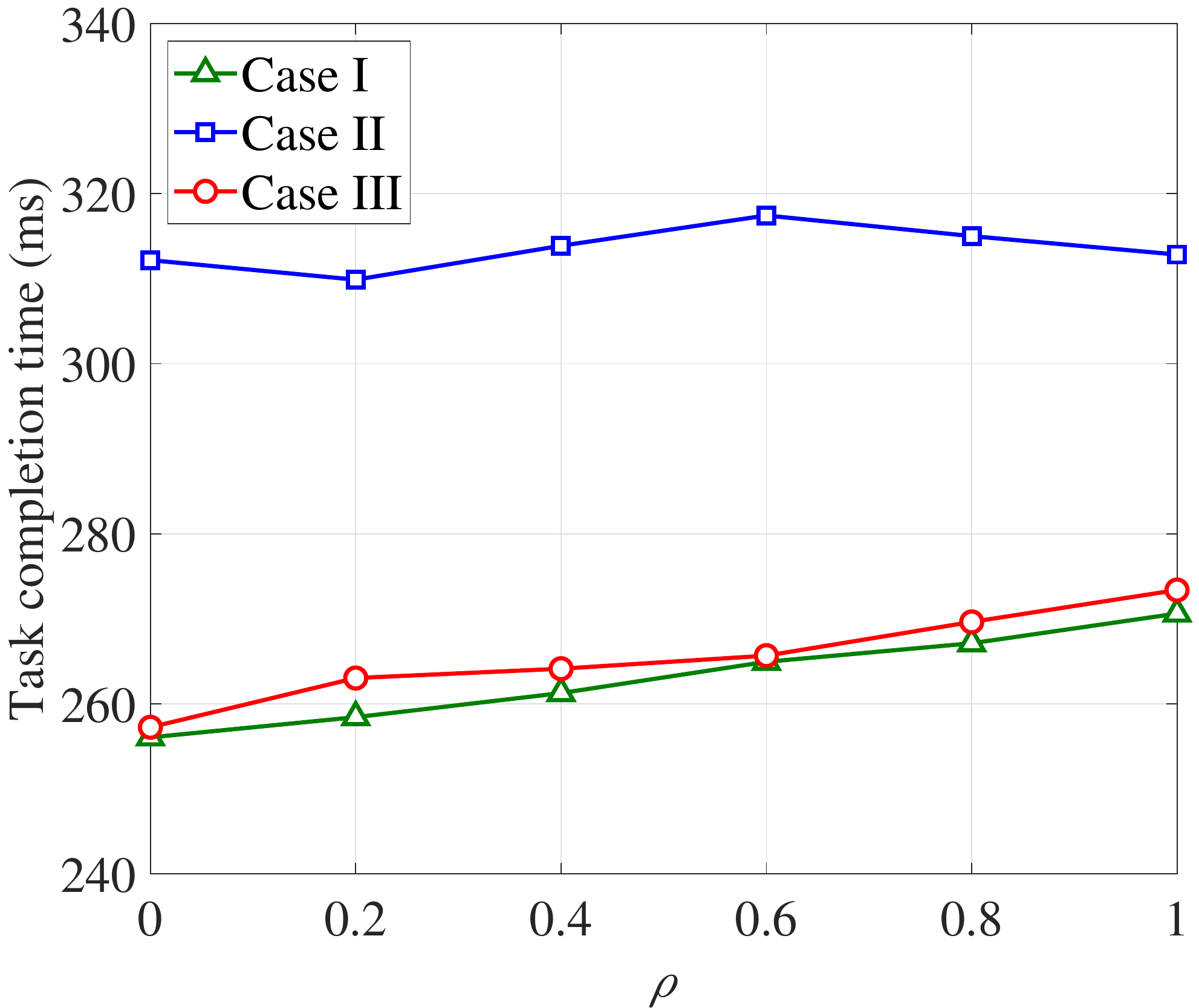}
		\label{Fig:Fig7b}}
	\hfil
	\subfloat[]{\includegraphics[width=3.25in]{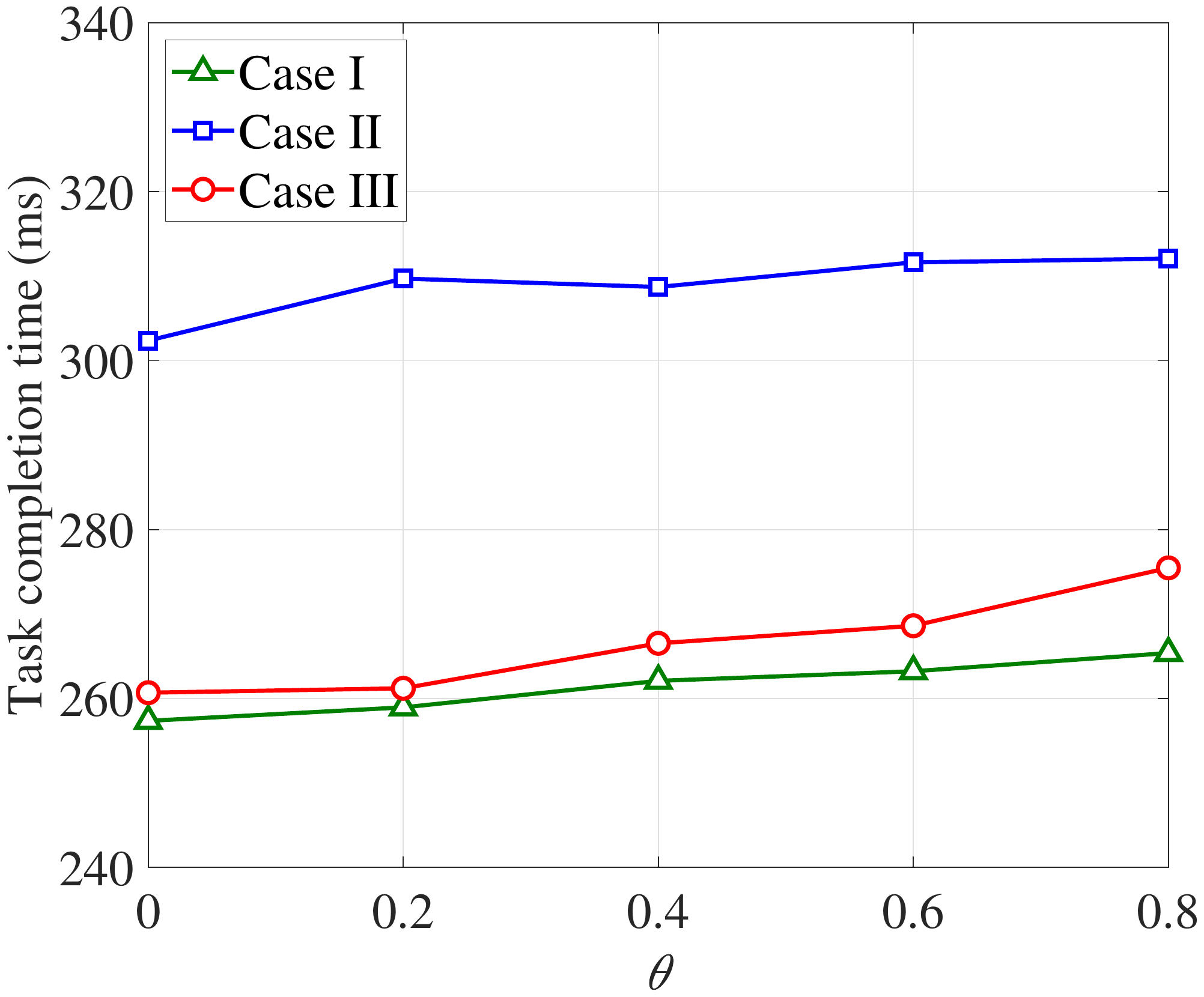}
		\label{Fig:Fig7c}}
	\caption{Task completion time comparison of three cases for joint velocity control and task offloading. 
		(a) $\rho=0.1$ and $\theta=0.1$. 
		(b) $N_{\rm CH}=4$ and $\theta=0.1$. 
		(c) $N_{\rm CH}=4$ and $\rho=0.1$.}
	\label{Fig:Fig7}
\end{figure}

In Fig. \ref{Fig:Fig7}\subref{Fig:Fig7a}, we first compare the task completion times of these three cases for different values of $N_{\rm CH}$, where $\rho=0.1$ and $\theta=0.1$.
We also plot the time consumption curves for different migration ratios in Fig. \ref{Fig:Fig7}\subref{Fig:Fig7b}, where $N_{\rm CH}=4$ and $\theta=0.1$.
Moreover, we plot the time consumption curves for different moving factors in Fig. \ref{Fig:Fig7}\subref{Fig:Fig7c}, where $N_{\rm CH}=4$ and $\rho=0.1$.
Observe from Fig. \ref{Fig:Fig7} that as the parameter values involving $N_{\rm CH}$, $\rho$, and $\theta$ increase, all task completion times increase. 
Furthermore, Case II has a much higher service delay than Cases I and III, and Case III has a slightly higher service delay than Case I.

\begin{figure}[tp]
	\centering
	\subfloat[]{\includegraphics[width=3.2in]{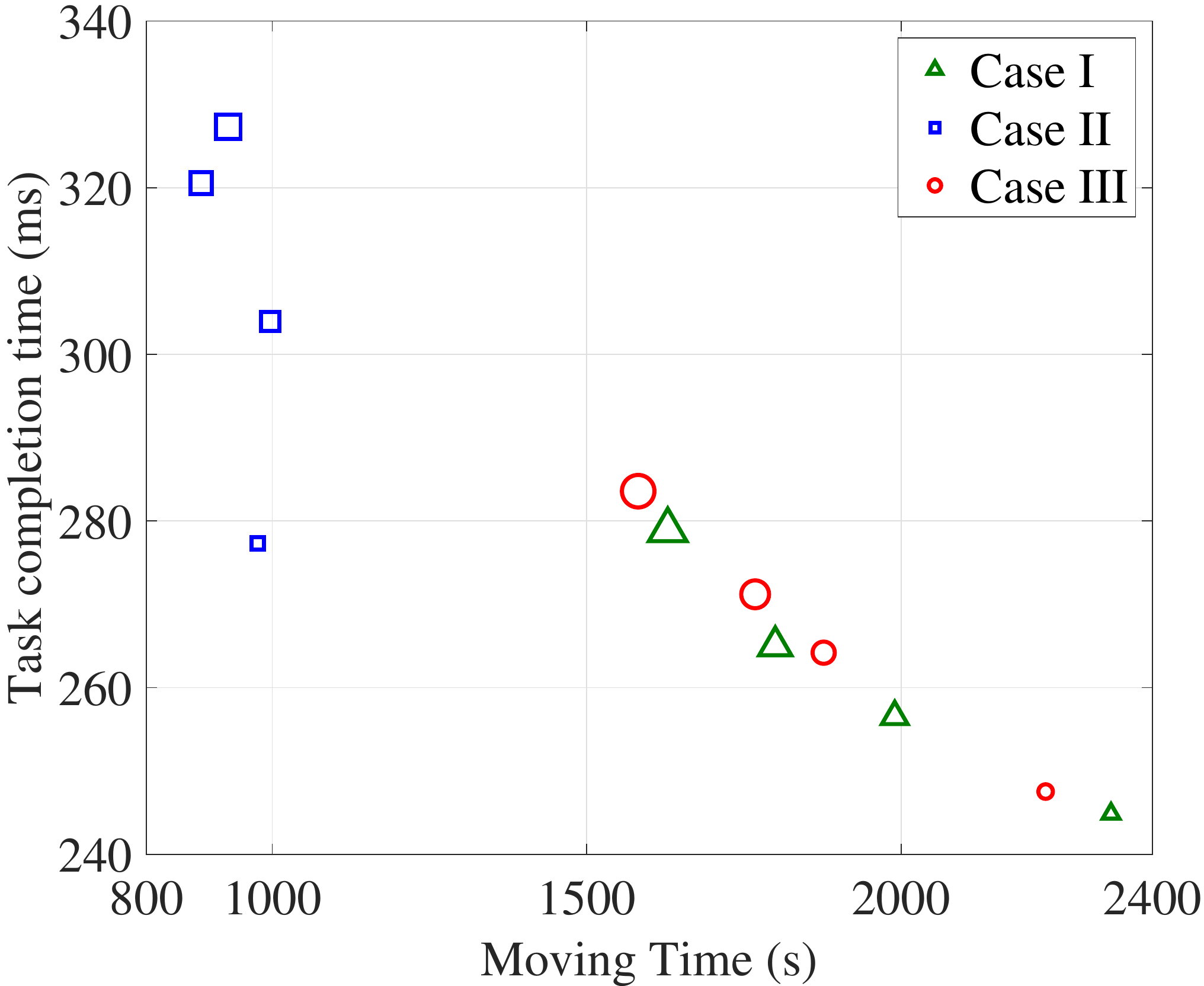}
		\label{Fig:Fig8a}}
	\hfil
	\subfloat[]{\includegraphics[width=3.2in]{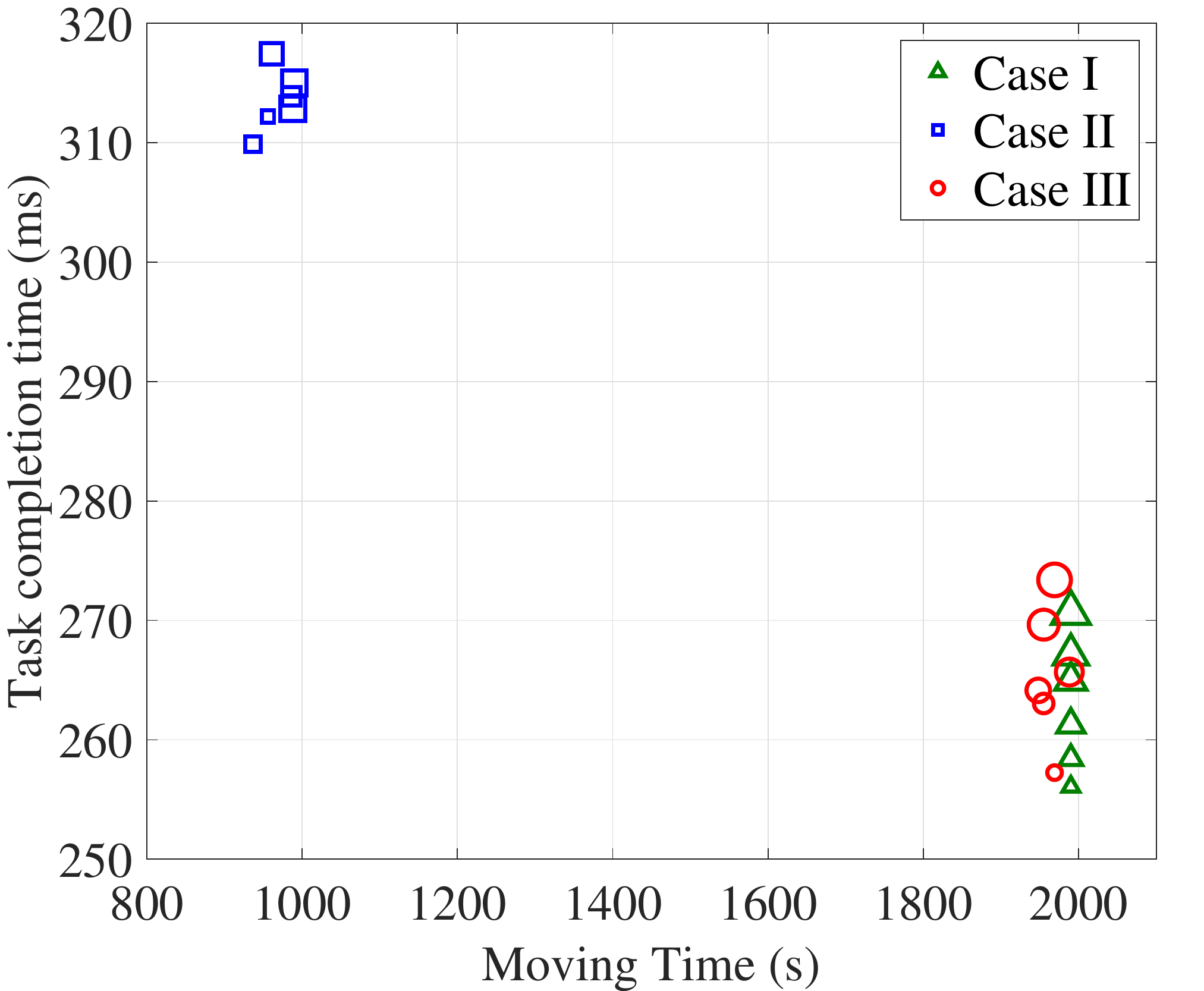}
		\label{Fig:Fig8b}}
	\hfil
	\subfloat[]{\includegraphics[width=3.2in]{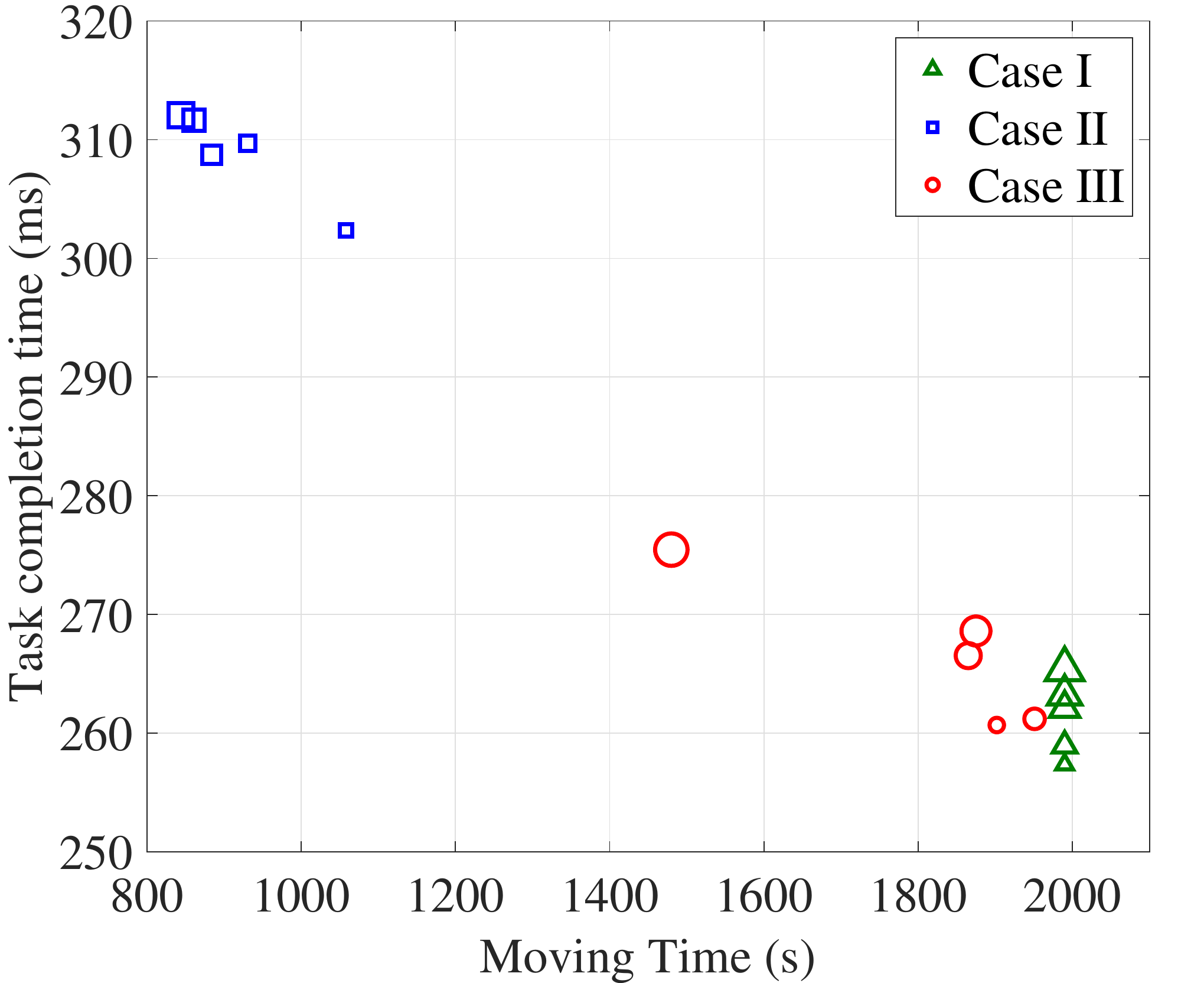}
		\label{Fig:Fig8c}}
	\caption{Task completion time versus moving time among three cases for joint velocity control and task offloading. 
		(a) $N_{\rm CH}=2,4,6,8$, $\rho=0.1$, and $\theta=0.1$. When the marker size increases, the value of $N_{\rm CH}$ increases from 2 to 8. 
		(b) $N_{\rm CH}=4$, $\rho=0,0.2,0.4,0.6,0.8,1$, and $\theta=0.1$. When the marker size increases, the value of $\rho$ increases from 0 to 1. 
		(c) $N_{\rm CH}=4$, $\rho=0.1$, and $\theta=0,0.2,0.4,0.6,0.8$. When the marker size increases, the value of $\theta$ increases from 0 to 0.8.}
	\label{Fig:Fig8}
\end{figure}

From Fig. \ref{Fig:Fig8}, the task completion time versus moving time performance is compared for Cases I, II, and III.
The parameters are set as follows: 
1) $N_{\rm CH}=2,4,6,8$, $\rho=0.1$, and $\theta=0.1$ in Fig. \ref{Fig:Fig8}\subref{Fig:Fig8a}; 
2) $N_{\rm CH}=4$, $\rho=0,0.2,0.4,0.6,0.8,1$, and $\theta=0.1$ in Fig. \ref{Fig:Fig8}\subref{Fig:Fig8b};
and 3) $N_{\rm CH}=4$, $\rho=0.1$, and $\theta=0,0.2,0.4,0.6,0.8$ in Fig. \ref{Fig:Fig8}\subref{Fig:Fig8c}.
As shown in Fig. \ref{Fig:Fig8}\subref{Fig:Fig8a}, although Case II has a lower moving time than Cases I and III, it has a much higher time completion time than Cases I and III.
Furthermore, Case III can obtain a lower moving time than Case I at the price of a slightly increased task completion time.
In addition, Figs. \ref{Fig:Fig8}\subref{Fig:Fig8b} and \ref{Fig:Fig8}\subref{Fig:Fig8c} show that due to the fixed number of APs with unavailable wireless communication, the velocity control of Case III lacks flexibility.  
As a result, Case III cannot adaptively reduce the moving time.
In contrast, Case III can dramatically reduce the moving time, especially for a large value of the moving factor shown in Fig. \ref{Fig:Fig8}\subref{Fig:Fig8c}, while maintaining a close task completion time to Case I.

In summary, the proposed scheme can create an elegant balance between service delay reduction and moving time reduction, and thus, it is a better choice for task offloading and velocity control compared to conventional schemes.

\section{Conclusion}

In this paper, a joint optimization problem of velocity control and task offloading has been proposed in a hybrid satellite-terrestrial network with multiple MEC servers.
To reduce the service delay for a mobile robot caused by increased local computations and frequent service migrations, the effect of wireless communication availability and velocity control on task offloading has been studied.
The analytical results of convergence rate and sample complexity of the improved $Q$-learning algorithm have been obtained.
Simulation results have shown that, unlike conventional counterparts, based on velocity control, the proposed scheme can obtain an effective offloading performance improvement in terms of the task completion time.
It was found that for mobile robots, mobility control is beneficial for providing high-quality service offloading in complex network environments.
However, further study is required to determine its effectiveness in the scenario of multiple mobile robots for multiple cooperative missions.
While this paper has only considered the decision-making of offloading and velocity control for one robot in the multi-robot environment, other forms of dynamics, such as time-varying bandwidth, dynamic computational resources, and complex trajectory planning, have considerable impact on MEC-based offloading.
These issues will constitute the direction of our future work.

\appendices
\section{Proof of Theorem \ref{Theo:Theo1}}
\label{App:App1}

	We first assume that the mobile robot has a constant velocity, that is, $\beta_{n}=0$, while the constraint $T_{{\rm goal},n} \leqslant k_n T_{\rm move}$ is satisfied.
	Thus, \eqref{eq:eq12} is simplified as
	\begin{subequations} \label{EQ:EQ1}
		\begin{align}
			\min\limits_{\textbf{1}_{\alpha}(m,t)} & T_{\rm mean} \label{EQ:EQ1a} \\
			\text{s.t.}~~
			& T_n(t) \leqslant T_{n,{\rm max}}(t) \label{EQ:EQ1b} \\
			&\sum\limits^{N}_{m=1}\textbf{1}_{\alpha}(m,t) \leqslant 1  \label{EQ:EQ1c} \\
			&\textbf{1}_{\alpha}(m,t) \in \{0,1\}  \label{EQ:EQ1d}
		\end{align}
	\end{subequations} 
	According to \cite{CATTRYSSE1992260}, the generalized assignment problem (GAP) can be formulated as
	\begin{subequations} \label{EQ:EQ2}
		\begin{align}
			\min\limits_{x_{ij}} & \sum\limits_{i}\sum\limits_{j} c_{ij} x_{ij} \label{EQ:EQ2a} \\
			\text{s.t.}~~
			& \sum\limits_{j} a_{ij} x_{ij} \leqslant b_i  \label{EQ:EQ2b} \\
			&\sum\limits_{i}  x_{ij} = 1  \label{EQ:EQ2c} \\
			&x_{ij} \in \{0,1\}  \label{EQ:EQ2d}
		\end{align}
	\end{subequations}
	where $c_{ij}$ is the cost of assigning Task $j$ to Agent $i$, $a_{ij}$ denotes the required capacity when Task $j$ is assigned to Agent $i$, and $b_i$ is the available capacity of Agent $i$.
	Based on \cite{CATTRYSSE1992260,bernhard2008combinatorial}, GAP is NP-hard.
	
	If we set $x_{ij}=\textbf{1}_{\alpha}(m,t)$, $b_i=\sum^{L_n}_{l=1} T_{n,{\rm max}}(l)$, $c_{ij}=T_{n}(t)/(\textbf{1}_{\alpha}(m,t) \sum^{N}_{n=1} L_{n})$, $\sum^{N}_{m=1}\textbf{1}_{\alpha}(m,t) = 1$, and $a_{ij}=T_{n}(t)/\textbf{1}_{\alpha}(m,t)$, GAP is a special case of \eqref{EQ:EQ1}. 
	Thus, \eqref{EQ:EQ1} is NP-hard.
	Furthermore, since \eqref{EQ:EQ1} is a special problem of \eqref{eq:eq12}, according to \cite{garey1978strong}, the optimization problem \eqref{eq:eq12} is NP-hard.

\section{Proof of Theorem \ref{Theo:Theo2}}
\label{App:App2}

	A constant-moving robot is first assumed, that is, $\beta_{n}=0$, and the constraint $T_{{\rm goal},n} \leqslant k_n T_{\rm move}$ should be satisfied. 
	Then, the coverage region decomposition-based optimization problem \eqref{eq:eq13} can be formulated as
	\begin{align}
		\max\limits_{\alpha_m(l)} & -\frac{1}{L_n} \sum\limits^{L_n}_{l=1} T_n(l) \label{EQ:EQ3} \\
		\text{s.t.}~~
		& \eqref{EQ:EQ1b}-\eqref{EQ:EQ1d}  \nonumber 
	\end{align}
	
	The 0-1 knapsack problem is expressed as follows.
	In a knapsack, its weight capacity is $W$.
	For a set of items, item $n$ has a weight of $w_n$ and a value of $v_n$.
	The objective is maximizing the summed value of items that can be packed in the knapsack, while maintaining the summed weight of items less than or equal to the weight capacity $W$.
	Thus, the 0-1 knapsack problem is formulated as
	\begin{subequations} \label{EQ:EQ4}
		\begin{align}
			\max\limits_{\mathbf{O}} &  \sum\limits_{n \in \mathbf{O}} v_n \label{EQ:EQ4a} \\
			\text{s.t.}~~
			& \mathbf{O} \subseteq \mathbf{I}  \label{EQ:EQ4b} \\
			& \sum\limits_{n \in \mathbf{O}} w_n \leqslant W  \label{EQ:EQ4c} 
		\end{align}
	\end{subequations}
	where $\mathbf{O}$ denotes the set of items that should be packed.
	According to \cite{bernhard2008combinatorial,8959146,Karp1972}, the 0-1 knapsack problem is NP-hard.
	
	If we set $v_n=-T_n/L_n$, $\mathbf{I}= \mathcal{L}_n$, $w_n=\alpha_m(l)$, and $W=1$, the 0-1 knapsack problem becomes a special case of \eqref{EQ:EQ3} for $T_n(t) \leqslant T_{n, {\rm max}}(t)$.
	Thus, the problem \eqref{EQ:EQ3} is also NP-hard.
	Moreover, since \eqref{EQ:EQ3} is a special case of \eqref{eq:eq13}, according to \cite{garey1978strong}, problem \eqref{eq:eq13} is also NP-hard.

\bibliographystyle{IEEEtran}
\bibliography{IEEEabrv,MyBibReference}

\end{document}